\documentclass[preprint,authoryear,11pt]{elsarticle}
\usepackage[letterpaper, left=1in, top=1in, right=1in, bottom=1in]{geometry}

\usepackage[utf8]{inputenc}
\usepackage[hang,flushmargin]{footmisc}
\usepackage[colorlinks,citecolor=blue,urlcolor=blue]{hyperref}
\usepackage{graphicx,algorithmic,amsmath,amssymb,rotating,parskip,xcolor,enumitem,natbib,charter,placeins,multirow,colortbl}

\usepackage[compact]{titlesec}
\titlespacing{\section}{0pt}{5pt}{3pt}
\titlespacing{\subsection}{0pt}{5pt}{3pt}
\titlespacing{\subsubsection}{0pt}{4pt}{3pt}
\hypersetup{
colorlinks,
citecolor=NavyBlue  
}
\bibpunct{(}{)}{;}{a}{}{,}

\def\para#1{\medskip\noindent{\bf #1}}  
\newcommand{\CR}{\cellcolor{red!25}}
\newcommand{\CG}{{\cellcolor{green!25}}}

\newcommand{\rot}[1]{\multicolumn{1}{c}{\rotatebox{90}{#1}}}
\newcommand{\red}{\color{red}}
\newcommand{\blue}{\color{blue}}

\def\eqn#1{eqn.~\eqref{#1}}
\def\twoeqns#1#2{eqns.~(\ref{#1},\ref{#2})}
\def\bx{\mathbf{x}}
\def\by{\mathbf{y}}
\def\bm{\mathbf{m}}
\def\ba{\mathbf{a}}

\def\be{\mathbf{e}}
\def\boldf{\mathbf{f}}
\def\bp{\mathbf{p}}
\def\bw{\mathbf{w}}
\def\bF{\mathbf{F}}
\def\bA{\mathbf{A}}
\def\bB{\mathbf{B}}
\def\bC{\mathbf{C}}
\def\bD{\mathbf{D}}

\def\bP{\mathbf{P}}
\def\bQ{\mathbf{Q}}
\def\bR{\mathbf{R}}
\def\bG{\mathbf{G}}
\def\bW{\mathbf{W}}
\def\bV{\mathbf{V}}
\def\bI{\mathbf{I}} 
\def\bzero{\mathbf{0}}
\def\bone{\mathbf{1}}
\def\bphi{\boldsymbol{\phi}}
\def\btheta{\boldsymbol{\theta}}
\def\bmu{\boldsymbol{\mu}}
\def\bnu{\boldsymbol{\nu}}
\def\bgamma{\boldsymbol{\gamma}}
\def\bomega{\boldsymbol{\omega}}
\def\bTheta{\boldsymbol{\Theta}}
\def\bGamma{\boldsymbol{\Gamma}}
\def\bLambda{\boldsymbol{\Lambda}}
\def\bSigma{\boldsymbol{\Sigma}}
\def\bOmega{\boldsymbol{\Omega}}

\def\cD{\mathcal{D}}

\def\tibm{\widetilde{\mathbf{m}}}
\def\tibC{\widetilde{\mathbf{C}}}
\def\tin{\widetilde{n}}
\def\tis{\widetilde{s}}
\def\tip{\widetilde{p}}

\begin{document}

\begin{frontmatter}

\title{Bayesian Forecasting \& Scalable Multivariate Volatility Analysis \\ 
 Using Simultaneous Graphical Dynamic Models \\ \quad \\ 
}

\author{Lutz F. Gruber$^1$}\ead{lutz.gruber@gmail.com}
\author{Mike West$^2$} \ead{mw@stat.duke.edu}
\address{\normalsize Duke University} 
\fntext[jn]{{\em Present position and address}: Senior Associate, QuantCo, Germany} 
\fntext[mw]{{\em Corresponding author}: The Arts \& Sciences Professor of Statistics \& Decision Sciences, Department of Statistical Science, Duke University, Durham 27708-0251, U.S.A. {Tel:} +1 919 684 8842}

\begin{abstract}
\noindent The recently introduced class of simultaneous graphical dynamic linear models (SGDLMs) defines an ability to scale on-line Bayesian analysis and forecasting to higher-dimensional time series. This paper advances the methodology of SGDLMs, developing and embedding a novel, adaptive method of simultaneous predictor selection in forward filtering for on-line learning and forecasting. The advances include developments in Bayesian computation for scalability, and a case study in exploring the resulting potential for improved short-term forecasting of large-scale volatility matrices.  A case study concerns financial forecasting and portfolio optimization with a 400-dimensional series of daily stock prices. Analysis shows that the SGDLM forecasts volatilities and co-volatilities well, making it ideally suited to contributing to quantitative investment strategies to improve portfolio returns. We also identify performance metrics linked to the sequential Bayesian filtering analysis  that turn out to define a leading indicator of increased financial market stresses, comparable to but leading the standard St. Louis Fed Financial Stress Index (STLFSI) measure.  Parallel computation 
using GPU implementations substantially advance the ability to fit and use these models. 
\end{abstract}

\begin{keyword}
Bayesian forecasting and portfolio optimization; dynamic graphical models; financial risk index;  GPU computation; high-dimensional time series; sparse multivariate stochastic volatility models. 
\end{keyword}

\end{frontmatter}
 
\newpage

\section{Introduction \label{sec:Intro}}
In time series portfolio analysis as in other areas of multivariate dynamic modeling and decision analysis 
in econometrics and finance, sparse models and efficient computation are critical to successfully scaling analyses to higher dimensional problems. With a focus on forecasting in financial time series, some of the recent progress with Bayesian sparsity modeling approaches---such as copula-based dynamic models~\citep[e.g.][]{gruberczado2015a}, 
dynamic graphical models~\citep[e.g.][]{CarvalhoWest2007, WangWest2009, QuintanaCarvalhoScottCostigliola2010, Wang2010} and sparse factor models~\citep[e.g.][]{West2003, YoshidaWest2010, CarvalhoLopesAguilar2011}---have been demonstrably useful. Forecasting improvements can be generated by data-relevant and informed sparsity, coupled with time-varying model parameters and relevant approaches to representing multivariate stochastic volatility~\citep[e.g.][]{quintanawest1987, harveyruizshephard1994, jacquierpolsonrossi2004, chankohnkirby2005, chibnardarishephard2006, LopesMcCullochTsay12}. Such advances can then be expected to aid in improved characterizations of risk and in outcomes of sequentially revised portfolio decision strategies~\citep[e.g.][]{PittShephard1999, AguilarWest2000, QuintanaLourdesAguilarLiu2003, ZhouNakajimaWest2012,ZhaoXieWest2015}. Examples in the above referenced papers and others in recent times typically involve series in just a few dimensions, although some show simulations and empirical results in up to 50 dimensions. For both institutional and personal implementations for quantitative investing and automated trading, and also in view of regulatory requirements on banks to assess market risk through value-at-risk or other metrics \cite[Paragraph 738]{BASEL2}, there is increasing interest in scaling methodology to substantially higher dimensions, at least to hundreds of series. 

Recently introduced simultaneous graphical dynamic linear models~\citep[SGDLMs:][]{gruberwest2015a} address scalability. These involve: (i) a sets of {\em decoupled} univariate 
dynamic linear models (DLMs) for individual series, allowing a range of time-varying parameter models and univariate volatilities, and for which standard theory and resulting efficient forward filtering/forecasting algorithms apply; (ii) exploitation of a simultaneous equations formulation with sparse graphical modeling ideas that {\em recouple}  the series and define rich yet sparse representations of multivariate stochastic volatility; and (iii) variational Bayes methods combined with importance sampling to coherently integrate/couple the series for forecasting and decisions. Parallel, GPU-based implementation enables on-line analysis of increasingly high-dimensional time series.

This paper defines and illustrates methodological advances in SGDLMs addressing core questions of variable selection underlying the dynamics in structure of large multivariate volatility matrices.  We develop and showcase this in a case study in financial forecasting and portfolio optimization with a 401-dimensional series of  daily S\&P 500 stock prices and index over 2003--2014.  The new methodology defines a strategy for 
sequential, adaptive selection of
simultaneous/contemporaneous parental predictor series of each index, and its use in the
case study highlights the utility in Bayesian forecasting and  portfolio decisions. The S\&P analysis includes benchmarks of forecast performance as well as portfolio returns and risk metrics, 
including comparisons to the 
standard multivariate Wishart DLM (WDLM)~\citep[][chap. 10]{PradoWest2010}. This is the appropriate benchmark as it has been a standard model in Bayesian financial time series and portfolio analysis---in industry and academic research---for years, being quite flexible and trivially implemented, and remains a mainstay 
component of many models.  

Section~\ref{sgdlm2:sec:forecast-models} introduces notation of DLMs, and briefly summarizes the key concepts and technical elements of the SGDLM. The SGDLM requires specification of a set of ``parental'' time series to use as contemporaneous predictors of each univariate series in a simultaneous equations formulation; to address this, Section~\ref{sgdlm2:sec:sgdlm-bayes-hotspot} introduces a novel and practicable selection strategy for the parental sets. Section~\ref{sgdlm2:sec:investment-strategy} discusses several quantitative investment rules based on various portfolio utility functions of practical interest. Section~\ref{sgdlm2:sec:data-example} presents a portfolio manager's view of managing a 400-asset portfolio using the SGDLM combined with such rules to drive investment decisions. Some summary comments  appear in Section~\ref{sgdlm2:sec:final-remarks}. Supporting technical material on WDLMs and SGDLMs, together with 
additional summaries from the case study, appear in the Appendix.   

\section{Forecasting Models} \label{sgdlm2:sec:forecast-models}

\subsection{Dynamic Linear Models (DLMs)} 

DLMs \citep{PradoWest2010, WestHarrison1997} are fully Bayesian state-space models that are widely used in forecasting financial time series due to flexibility in model specification, ability to adapt to changing market dynamics and to incorporate external/intervention information.
The standard univariate DLM combines a normal linear observation equation,
\begin{equation} \label{sgdlm2:eq:univariate-DLM-obs}
y_t = \bF_t \btheta_t + \nu_t \text{,}
\end{equation}
with a conditionally normal, multivariate linear system equation to govern the state evolutions of $\btheta_t$ from time $t$ to $t+1$,
\begin{equation} \label{sgdlm2:eq:univariate-DLM-system}
\btheta_{t+1} = \bG_{t+1} \btheta_t + \bomega_{t+1} \text{.}
\end{equation}
Here the observation errors $\nu_t \sim N(0, \lambda_t^{-1})$ follow a normal distribution with precision $\lambda_t$, and the state innovations $\bOmega_t \sim N(\bzero, \bW_t)$ are multivariate normally distributed with covariance $\bW_t$. 
In financial time series,  the necessity of volatility models is well-understood, and standard extensions of the
basic DLM to include the beta-gamma stochastic evolution of the precisions $\lambda_t$ are widely used;
see key source and references in~\citet[][sect. 10.8]{WestHarrison1997} and 
\citet[][sect. 4.3]{PradoWest2010}. Details applied to the SGDLM are elaborated in the following section. 
Conjugate analysis enables fast, on-line learning, so that models are updated dynamically, responding to the latest market events, while being open to user-intervention at all times.

The widely-used, benchmark  multivariate DLMs with dynamic volatility matrices extend the above univariate
model to a vector time series in which the variance matrix of observation errors evolves according to 
a multivariate beta-Wishart process. Again theory is standard; see~\citet[][sect. 16.4]{WestHarrison1997} and 
\citet[][chap. 10]{PradoWest2010}. We denote this model by WDLM, and give key summary details below in \ref{sgdlm2:subsec:wdlm}.

\subsection{SGDLM} \label{sgdlm2:subsec:sgdlm}

The SGDLM combines univariate DLMs for each series to define a multivariate model, and does this
via contemporaneous regressions of each series on a subset of the other series. This  allows for parsimonious modeling of multivariate dependence for enhanced scalability, and was recently introduced   in~\cite{gruberwest2015a}. We summarize the essential details here.

\para{Observation model.} 
Each of the $m$ univariate series $y_{jt}$, $j=1{:}m$, follows a univariate DLM with observation equation 
\begin{equation}
y_{jt} = \bF_{jt}' \btheta_{jt} + \nu_{jt} = \bx_{jt}' \bphi_{jt} + \by_{sp_t(j),t}'\bgamma_{jt} + \nu_{jt}, \qquad \nu_{jt} \sim N(0, 1/\lambda_{jt}), \label{sgdlm2:eq:SGDLMj} \\
\end{equation}
where: 
\begin{itemize} 
\item  The predictor vector $\bF_{jt} = (\bx_{jt}, \by_{sp_t(j),t})'$ consists of $p_{j,\phi}$ external predictors $\bx_{jt} $ to model the local level of $y_{jt}$, together with the values of $p_{j,\gamma}$ {\em contemporaneous} series $\by_{sp_t(j),t}$ with indices in the {\em simultaneous parental set} $sp_t(j) \subseteq \{1,\ldots,m\} \setminus \{j\}$ of size $|sp_t(j)| = p_{j,\gamma}$. The latter allows for effective modeling of cross-series, time-varying conditional dependencies and across $j$ this defines a simultaneous equations formulation of multivariate volatility. 
\item The state vector $\btheta_{jt} = (\bphi_{jt}, \bgamma_{jt})'$ is partitioned accordingly: $\bphi_{jt}$ is the $p_{j,\phi}$-regression vector of $\bx_{jt}$ and $\bgamma_{jt}$ that for the $p_{j,\gamma}$ simultaneous parental coefficients $\by_{sp_t(j),t}$. 
\item The precision process $\lambda_{jt}$ allows modeling of residual stochastic volatility patterns over time. 
\item Conditional on the volatility processes,  the zero-mean normal observation errors $\nu_{jt}$ are independent across series $j$ and time $t.$
\end{itemize} 
Write $\bmu_t = (\mu_{1t}, \ldots, \mu_{mt})'$, where $\mu_{jt} = \bx_{jt}'\bphi_{jt}$, and $\bLambda_t = (\lambda_{1t}, \ldots, \lambda_{mt})$. Furthermore, write $\bGamma_t = (\gamma_{j=1{:}m,h=1{:}m,t})$ for the matrix that contains the elements of the simultaneous parental coefficients $\bgamma_{jt}$, with extension to $\gamma_{j,h,t} = 0$ if $h \not \in sp_t(j)$. Conditional on these quantities, the multivariate series is conditionally normal, 
\begin{equation} \label{sgdlm2:eq:SGDLM-y-mvn}
\by_t \sim N(\bA_t \bmu_t, \bSigma_t) \text{,}
\end{equation}
where $\bA_t = (\bI - \bGamma_t)^{-1}$ and $\bOmega_t \equiv \bSigma_t^{-1} = (\bI - \bGamma_t)' \bLambda_t (\bI - \bGamma_t)$. The SGDLM allows for modeling flexibility in that different external predictors $\bx_{jt}$ can be selected for each series. Furthermore, the simultaneous parental specification of the volatility matrix $\bSigma_t$ allows for sparse models since the sizes $p_{j,\gamma}$ of the parental sets $sp_t(j)$ can be chosen much smaller than $m$. 

The states and precisions evolve according to a standard DLM~\citep[][chap. 4]{PradoWest2010}  with a linear, Gaussian state evolution for $\btheta_{jt}$  coupled to a discount volatility model for $\lambda_{jt},$ enabling closed-form computations for sequential filtering and forecasting. Full specification involves cumulated information summarized in conditionally conjugate   distributions, as follows. 

\para{Priors at time $t$.} Independently across series, the prior for the series $j$ state vector and precision is normal/gamma 
\begin{equation}
 (\btheta_{jt},\lambda_{jt}|\cD_{t-1}) \sim NG(\ba_{jt},\bR_{jt},r_{jt},s_{j,t-1}), \qquad j=1{:}m. \label{sgdlm2:eq:timet-1prior}
\end{equation}
In this $NG$ notation, $\btheta_{jt}|\lambda_{jt} \sim N(\ba_{jt},\bR_{jt}/(s_{j,t-1}\lambda_{jt}))$ and $\lambda_{jt} \sim G(r_{jt}/2,r_{jt}s_{j,t-1}/2)$ with shape $r_{jt}/2>0$, rate $r_{jt}s_{j,t-1}/2 > 0$ and mean $1/s_{j,t-1}$.  The implied $\btheta_{jt}$ margin is multivariate T with $r_{jt}$ degrees of freedom, mode $\ba_{jt}$ and scale matrix $\bR_{jt}$.

With $\bTheta_t= [ \btheta_{1t},\ldots,\btheta_{mt}]$ and  $\bLambda_t = (\lambda_{1t}, \ldots, \lambda_{mt})$ the joint prior across series has density 
\begin{equation}
p(\bTheta_{t}, \bLambda_{t} | \cD_{t-1}) = \prod_{j=1{:}m} p_{jt}(\btheta_{jt}, \lambda_{jt}| \cD_{t-1}). \label{sgdlm2:eq:sgdlm-prior-t}
\end{equation}

\para{State evolution model.}   From $t-1$ to $t,$ the state evolves conditional on $\lambda_{jt}$ and $\cD_{t-1}$ via
\begin{equation}
\btheta_{jt} = \bG_{jt} \btheta_{j,t-1} + \bomega_{jt}, \qquad \bomega_{jt} \sim N(\bzero, \bW_{jt}/(s_{j,t-1}\lambda_{jt})), \label{sgdlm2:eq:SGDLM-state-evo2} 
\end{equation}
based on evolution matrices $\bG_{jt}$ and innovations $\bomega_{jt}$ having conditional variance matrices $\bW_{jt}$ scaled by $s_{j,t-1}\lambda_{jt}$.   Conditional on the model states, volatility processes, evolution transition and variance matrices, the zero-mean observation errors $\nu_{jt}$ and state innovations $\bomega_{jt}$ are independent and mutually independent across series $j$ and over time $t$.   The $\bW_{jt}$ matrices are specified using discount factors~\citep[][chap.~6]{WestHarrison1997} as detailed further below. 

\para{Forecasts at time $t$.} The one-step ahead predictive distributions are efficiently evaluated by simulation. Draw from the set of $m$ independent normal/gamma priors of~\eqn{sgdlm2:eq:timet-1prior} to define a simulation sample $\{ \bTheta_t^r, \bLambda_t^r \}$, where $r=1{:}R$ indexes Monte Carlo samples for prediction. Each sample defines Monte Carlo values of one-step forecast moments $\bA_t^r\bmu_t^r, \bSigma_t^r$, which can be used to simulate from the predictive distribution of $\by_t$ using the conditionally normal form of \eqn{sgdlm2:eq:SGDLM-y-mvn}.  Step-ahead forecasting more than one period is similarly easily done via simulation. 

\para{Recoupling the posterior at time $t$.} The exact posterior is
\begin{equation}\label{sgdlm2:eq:timetposterior}
p(\bTheta_t,\bLambda_t | \cD_t) \propto |\bI - \bGamma_t| \prod_{j=1{:}m} \tip_{jt}(\btheta_{jt},\lambda_{jt} | \cD_t) \text{,}
\end{equation}
where each $\tip_{jt}(\btheta_{jt},\lambda_{jt} | \cD_t)$ factor is of a normal/gamma form $NG(\tibm_{jt},\tibC_{jt},\tin_{jt},\tis_{jt})$ that arises from standard analytic updating of each series individually. The parameters are obtained as $\tibm_{jt} = \ba_{jt} + \bA_{jt} e_{jt}$, $\tibC_{jt} = (\bR_{jt} - \bA_{jt} \bA_{jt}' q_{jt}) z_{jt}$, $\tin_{jt} = r_{jt} + 1$ and $\tis_{jt} = z_{jt} s_{j,t-1}$, after first computing the forecast error $e_{jt} = y_{jt} - \bF_{jt}' \ba_{jt}$, forecast variance factor $q_{jt} = s_{j,t-1} + \bF_{jt}' \bR_{jt} \bF_{jt}$, adaptive coefficient vector $\bA_{jt} = \bR_{jt} \bF_{jt} / q_{jt}$ and volatility update factor $z_{jt} = (r_{jt} + e_{jt}^2/q_{jt})/(r_{jt} + 1)$. The determinant term appearing in the exact posterior above theoretically {\em recouples} the prior-independent states to account for between-series dependence effects. \cite{gruberwest2015a} show the efficacy of importance sampling to evaluate characteristics of the exact posterior of \eqn{sgdlm2:eq:timetposterior}. Samples from the independent normal/gammas $\tip_{jt}$ are importance-weighted by the resulting values of the determinant term $|\bI - \bGamma_t|$ to define the Monte Carlo approximation to the full joint posterior. 

\para{Decoupling for evolution to time $t+1$.} To enable independent parallel processing of prior evolutions across series $j$, the exact posterior is {\em decoupled} into a product of conjugate forms across the series $j=1{:}m$. This uses  a standard  mean-field variational Bayes (VB) approach~(e.g., \citealp[][sect. 12.3]{WestHarrison1997}; \citealp{JaakolaJordan2000}) that emulates the exact posterior by a product of normal/gammas
\begin{equation}\label{sgdlm2:eq:VBapproxpostt} 
q(\bTheta_t, \bLambda_t | \cD_t) = \prod_{j=1{:}m} q_{jt}(\btheta_{jt}, \lambda_{jt} | \cD_t) \ \ \textrm{with}\ \
(\btheta_{jt}, \lambda_{jt} | \cD_t) \sim NG(\bm_{jt}, \bC_{jt}, n_{jt}, s_{jt}), \ \ j=1{:}m.
\end{equation}
The VB strategy selects the parameters in~\eqn{sgdlm2:eq:VBapproxpostt} to minimize the Kullback-Leibler divergence of the product form $q(\cdot|\cdot)$ from the exact joint posterior $p(\cdot | \cdot)$.
\ref{sgdlm2:app:vb-decoupling} gives summary equations. 

\para{Evolution to time $t+1$.} Moving ahead one time point,  states and volatilities undergo evolutions. For each  $j$, the $\lambda_{jt}$ first evolves to $\lambda_{j,t+1}$ according to the standard gamma/beta stochastic volatility model; see, for example,~\citet[][sect. 10.8]{WestHarrison1997},~\citet[][sect. 4.3]{PradoWest2010}. This is based on a series-specific discount factor $\beta_j \in (0,1)$, typically close to 1. Following this, the state vector $\btheta_{jt}$ evolves to $\btheta_{j,t+1}$ according to the state evolution of~\eqn{sgdlm2:eq:SGDLM-state-evo2} but with $t$ updated to $t+1.$ The specification is such that the evolved priors at time $t+1$ maintain the normal/gamma form, enabling fast, exact analysis; resulting priors are precisely as in~\twoeqns{sgdlm2:eq:timet-1prior}{sgdlm2:eq:sgdlm-prior-t} with  $t$ updated to $t+1$. The parameter evolutions $\ba_{j,t+1} = \bG_{j,t+1} \bm_{jt}$, $\bR_{j,t+1} = \bG_{j,t+1} \bC_{jt} \bG_{j,t+1}' + \bW_{j,t+1}$ and $r_{j,t+1} = \beta_j n_{jt}$ follow standard DLM theory and notation as in the above references. 

Model completion requires specification of the $\bG_\ast$ and $\bW_\ast$ matrices. In the case study of Section~\ref{sgdlm2:sec:data-example}, each $\bG_\ast=\bI$ and we use block discounting to specify the $\bW_\ast$~\citep[][sect. 6.3]{WestHarrison1997}. For series $j,$ this uses two discount factors: $\delta_{j\phi}$ for the external predictor state vector, and  $\delta_{j\gamma}$ for the parental state vector,
with values satisfying $0\ll \delta_*<1.$ With $\bB_{j,t+1} = \bG_{j,t+1} \bC_{jt} \bG_{j,t+1}'$, this defines $\bW_{j,t+1}$ as a partitioned matrix with upper-left block diagonal $\bB_{j,t+1,\phi}(1/\delta_{j\phi} - 1)$,  lower-right block diagonal $\bB_{j,t+1,\gamma}(1/\delta_{j\gamma} - 1)$, and upper-right (covariance) block $\bB_{j,t+1,\gamma \phi}(1/\sqrt{\delta_{j\phi}\delta_{j\gamma}} - 1)$. 

\para{Computation.} Recoupling of the posteriors using importance sampling is the only computationally demanding step. This is well-suited to GPU-based massive parallelization since the $m$ model simulations are decoupled. As shown in~\cite{gruberwest2015a}, this makes fully Bayesian, real-time analysis with $m$ in the hundreds to thousands feasible. On standard 2014 GPU-enabled desktop machines, one full iteration of evolution/forecasting/updating takes less than 10 seconds with $m \approx 400$ and modest dimensional models. The software discussed in that reference is used here. 

\section{Forward Filtering Selection of Simultaneous Parental Sets} \label{sgdlm2:sec:sgdlm-bayes-hotspot}

\subsection{Perspective}

We will typically have $|sp_t(j)|$ much smaller than $m$ in problems where $m$ is at all large. With $m=401$ in our S\&P case study (Section~\ref{sgdlm2:sec:data-example}), there are many patterns of time-varying dependencies among stocks, but it is inappropriate to expect real practical value in estimating co-volatilities from models with more than, say, 20 or so simultaneous predictors. That is, the implied dynamic graphical model---represented by zeros/non-zeros in $\bGamma_t$ and $\bOmega_t$---will typically be quite sparse. Collinearities among potential simultaneous parental series will typically mean that many possible choices of a (smallish) parental set for any one series will yield similar predictions, so working with one set of selected $sp_t(j)$ 
over short time periods is desirable.


The perspective here is critical: we are not interested in formal inference on 
parental sets, and such sets will not typically be stable over time or practically identifiable in 
problems with many series.   Choices of 
parental sets are only interesting as vehicles to improved forecasts and decisions.   In larger problems, 
any choice of a set of, say 10 parents for one series for a particular short time period will be practically indistinguishable
from multiple other candidate sets in which some of the parents are replaced by strongly collinear alternatives.  
Rather, the perspective is to identify small parental sets and 
adaptively revise them over time to capture and characterize the structure and dynamics of resulting 
(precision and variance) volatility matrices.  Our goals and interests are forecasting and portfolio decisions, and the 
dynamic precision matrices drive core aspects of the overall Bayesian decision analysis. 
Hence, we define a novel strategy to systematically and adaptively select/revise the $sp_t(j)$ over time,
applied separately-- in parallel-- to each series.

\subsection{Forward Filtering Selection: Concept}

Standing at a given time $t,$  partition each set $sp_t(j)$ into three categories: a dynamic {\em \lq\lq core\rq\rq} group of  
simultaneous parents, $sp_{\text{core},t}(j)$; a set of candidate simultaneous parents $sp_{\text{up},t}(j)$; and a set of outgoing simultaneous parents $sp_{\text{down},t}(j)$. The core simultaneous parents define the
current sparsity structure of the SGDLM's and underlie cross-series links in terms of precisions/co-volatilities. The warm-up groups serve to inform learning on the dynamic posterior distribution of the simultaneous regression coefficients $\gamma_{sp_{\text{up},t}(j),jt}$ of recently added simultaneous parental series in combination with the existing simultaneous parents as the full SGDLM is filtered forward. The outgoing group contains parents that are eliminated from either the warm-up group or core group, and are phased-out over several time steps by gradually shrinking their  coefficients $\gamma_{sp_{\text{down},t}(j),jt}$ to zero. 

The dimensions of the three simultaneous parental sets are defined by the modeller. While we see merit in the use of approaches such as dynamic latent thresholding~\citep[e.g.][]{NakajimaWest2010,ZhouNakajimaWest2012} that set these dimensions autonomously, such approaches are simply not adaptable to forward filtering and forecasting contexts with 
higher dimensional series. Computational issues are a barrier, but -- more importantly-- the perspective that 
we care only about useful predictive models, and not at all about specific parental sets that may be playing roles, 
indicates that such approaches are overkill.     We focus on a more direct selection strategy that is
consistent with this perspective and that provides an elegant solution to several typical problems of dynamic model selection.
Our Bayesian strategy has a number of practically key features, now noted and then elaborated in following discussion. 
Specifically:
\begin{itemize}
\item Forward filtering selection requires very little additional computation time, and is substantially better-suited for on-line application than conducting any kind of formal Bayesian model search at every step $t$. Methods based on Markov chain, sequential Monte Carlo algorithms or related stochastic search methods are simply infeasible (technically and) computationally, as well as philosophically directed towards goals that are not relevant in our
contexts.
\item The idea of a warm-up period to phase in new simultaneous parents uses data-informed posterior learning (over several time steps) and eliminates the need for delicate specification of initial priors of new simultaneous regression coefficients.
\item Phasing-in new simultaneous parents from neutral zero-mean initial priors, and phasing out existing simultaneous parents by gradual shrinkage to zero makes resulting forecasts of multivariate volatility patterns robust by, in part, 
inducing \lq\lq smooth" changes in the  structure of resulting dynamic predictive precision matrices. 
\end{itemize}

\subsection{Forward Filtering Selection: Strategy and Implementation}

\subsubsection{General strategy} 

At each $t$, we allow for changes in the \lq\lq current" parental set for each series. 
A key part of this is that candidate simultaneous parents lie in the warm-up sets $sp_{\text{up},t}(j)$.  
The size of each of these sets is a fixed value  $\Delta t = |sp_{\text{up},t}(j)|$; this value also
determines how many time steps each simultaneous parent is granted before it will either be included in  $sp_{\text{core},t  + \Delta t}(j)$, or be gradually eliminated via assignment to 
$sp_{\text{down},t  + \Delta t}(j)$.
 
Linking to formal and MCMC-based Bayesian variable selection, note that 
MCMC sampling consists of repeated performance of a proposal step and an acceptance/rejection step
for candidate variables to include or exclude.  Our forward filtering selection builds on this underlying MCMC concept, 
adapting it to the forward/sequential analysis with our explicit decision focus.  
The first  modification is that forward filtering selection uses only one proposal at each time point $t$, while MCMC sampling generates as many proposals as there are MCMC iterations. In our parental selection selection, the decision to accept or reject the time $t$ proposal is delayed to time $t + \Delta t$, at which point the proposed additional parents have been tentatively included in the $sp_{\text{up},t:t + \Delta t}(j)$, $j \in \{1{:}m\}$, since time $t$. By separating proposal generation from the acceptance decision, posterior information from joint updates of the regression coefficients of the proposed and existing simultaneous parents can be factored into the decision, and the choice of initial priors for newly added simultaneous parents becomes mostly irrelevant. Then, the acceptance rule is different from the typical Metropolis--Hastings acceptance probability, as the goal is to select between two alternatives and not to estimate posterior probabilities of every possible model specification.

\subsubsection{Strategy: Adding new simultaneous parents}
Our new strategy adaptively revises simultaneous parental sets based on a parallel analysis of the data using a standard WDLM. While this standard analysis is limited in terms of scalability and in its potential to predict changes in multivariate volatility patterns, it is able to track and adapt to such changes, so providing an obvious \lq\lq proposal" model for generating insights into parental structure.  The simple conjugate/analytic sequential analysis of the WDLM tracks and estimates the  $m\times m$ time-varying precision matrix $\bOmega_t$ without  constraints.  Inference on $\bOmega_t$ allows interrogation of resulting posterior Wishart distributions as they evolve over time. At any time $t,$  off-diagonal elements in row $j$  define conditional regression coefficients of all $m-1$ series $i\ne j$ in predicting $y_{jt}$. Larger absolute values of the precision elements in row $j$ thus suggest candidates for inclusion in the parental set $sp_t(j)$. In our case study, 
we consider series $k\ne j$ for inclusion in $sp_t(j)$ if the absolute value of the $(j,k)$ precision element is among the largest $n_{\text{max}}=10$ values in row $j$.  Each such series $k$ not already in the 
warm-up or core parental sets becomes a candidate in the warm-up set; i.e., each series $k$ among 
these \lq\lq top" $n_{\text{max}}$ 
is added to $sp_{\text{up},t}(j)$ if $k\not \in sp_{\text{up},t-1}(j) \cup sp_{\text{core},t-1}(j)$. 

This inclusion of series $k$ as a   \lq\lq candidate" parent of series $j$  involves specifying prior (at the current time $t$) information 
for the corresponding  coefficient $\gamma_{kjt};$ we take this as having  zero mean and a specified variance. Once the candidate parental series is embedded in the model, posterior information on its contribution and relevance  is generated during the regular evolution and updates over times $t+1,  \ldots, t+\Delta t.$  
After this learning period, the candidate parent is promoted into the core set $sp_{\text{core},t + \Delta t}(j)$. If this addition grows the core set beyond its target size, the additional as well as incumbent parental series 
are reviewed to drop
(or \lq\lq retire") one or more parents, as follows.

\subsubsection{Strategy: Dropping simultaneous parents}
At each time $t$ and for each series $j,$ simultaneous parents are retired from $sp_{\text{core},t}(j)$ if this set exceeds its target size through the addition of a new parents. This process  involves two  steps: the selection of the series that will be dropped, and then the phasing out of the selected series.
We target series $k$ for dropping based on inference on current values of parental coefficients $\gamma_{kjt}$, using
standardized posterior values  (a.k.a. signal-to-noise ratios) 
$
\text{SNR}_{kjt} = \ba_{kjt}/\bR_{kkjt}.
$
Parental predictors with small values of these ratios are candidates for the retirement, i.e., for 
inclusion in the phase-out set $sp_{\text{down},t}(j)$.  Elimination of regression effects with small signal-to-noise 
ratios is conducive to improving forecast performance and reliability. 

The simultaneous coefficients $\gamma_{sp_{\text{down},t}(j),jt}$ of outgoing  parents in $sp_{\text{down},t}(j)$ are gradually shrunk to zero over the next $\Delta t$ time steps. Shrinkage is implemented as prior intervention through the state evolution matrices $\bG_{jt}$ by appropriately scaling the corresponding diagonal entries as
\begin{equation} \label{sgdlm2:eq:evo-shrinkage-phase-out}
\bG_{kj,t_0 + l} = 1 - \{(\Delta t + 1) - l\}^{-1} \quad\text{for}\,\, l=1{:}\Delta t \text{,}
\end{equation}
relative to that time $t_0$ at which series $k$ was added to $sp_{\text{down},t_0}(j)$. The sequential shrinkage in~\eqn{sgdlm2:eq:evo-shrinkage-phase-out} results in stochastic reduction of the role of series $k$ to zero in $\Delta t$ steps. Shrinkage over several time steps allows for the roles of other simultaneous parents to adjust, and makes forecasts of the precision and covariance matrices of $\by_t$ more robust via the resulting \lq\lq smooth" transitions of parental predictors included/excluded. 
 
\section{Bayesian Portfolio Analysis} \label{sgdlm2:sec:investment-strategy}

Section~\ref{sgdlm2:sec:data-example} involves assessments of a range of dynamically optimized and updated portfolios 
based on Bayesian decision analysis under several portfolio utility functions~\citep[][sect. 10.4.7]{markowitz1952, markowitz1959, AguilarWest2000, CarvalhoWest2007, QuintanaLourdesAguilarLiu2003, QuintanaCarvalhoScottCostigliola2010, PradoWest2010}. We explore portfolio utilities that represent currently topical and relevant approaches in modern
quantitative investment management, all being extensions of traditional penalized mean-variance decision rules.  The analysis models daily log-returns on stocks and sequentially updates the portfolio allocation across these stocks via Bayesian decision analysis using chosen portfolio utility functions. Mean-variance optimization aims to control risk while aiming for positive returns, and modified utilities overlay additional, practically relevant constraints. In addition to specific {\em target return} portfolios, we consider utility functions that incorporate a benchmark index and require that optimized portfolios be, in expectation, uncorrelated with the benchmark in addition to target return and risk components. 

Our models are applied to the vector of daily log-returns $\by_t$. In all models, the mean and variance matrix of the one-step ahead forecast distribution $p(\by_t|\cD_{t-1})$ are key ingredients. Denote these by $\bp_t = E(\by_t|\cD_{t-1})$ and $\bP_t = V(\by_t|\cD_{t-1})$. 
In the SGDLM, these are computed via Monte Carlo simulation. A portfolio weight vector $\bw_t = (w_{1,t}, \ldots, w_{m,t})'$ defines the allocation of capital across the $m$ assets. The decision is to choose $\bw_t$ at market close on day $t-1$, and then act on that reallocation; on day $t$, the new closing prices are realized and the process repeats on the following day. Based on the forecast distribution of log-returns, the implied one-step ahead forecast mean and variance of the portfolio for any specific weight vector $\bw_t$ are $\bw_t'\bp_t$ and $\bw_t'\bP_t\bw_t$, respectively. 

\para{Minimum variance portfolio.} The standard or baseline minimum variance portfolio chooses $\bw_t$ as that vector minimizing the expected portfolio variance $\bw_t \bP_t \bw_t$ subject to $\bone'\bw_t = 1$. The optimal weight vector is trivially computed. More practically relevant portfolio strategies overlay additional constraints, as follows. 

\para{Target return mean-variance portfolio.} The original~\citep{markowitz1952, markowitz1959} mean-variance portfolio rule minimizes the risk---again in terms of portfolio variance---for a given, desired target return $\tau_t$. The relevant decision analysis simply modifies the minimum variance portfolio optimization by adding the constraint $\bw_t' \bp_t \geq \tau_t$, or its practical equivalent $\bw_t' \bp_t = \tau_t$. Note that the targets $\tau_t$ can vary over time, and be chosen adaptively by either direct specification or an automated rule.


\para{Benchmark-neutral portfolio.} This refinement mandates that the portfolio be uncorrelated, in expectation, at each step with a selected benchmark time series. To implement this, joint 1-step ahead forecast distributions are required for the assets of interest together with the benchmark series. With no loss of generality, we do this by taking the selected benchmark series as $j=1$. The relevant decision analysis then simply modifies the portfolio optimization rules above by adding the constraints $w_{1t}= 0$ and $\bw_t' \bP_{\cdot 1t}= \bzero$, where $\bP_{\cdot 1t}$ is the first column of $\bP_t$ containing the covariances of all series with the benchmark.

\section{Case Study: S\&P 500 Company Stocks} \label{sgdlm2:sec:data-example}
 
\subsection{Context and Data}
We use data on the S\&P 500 stock market index (SPX) and 400 S\&P 500 member stocks that have been continuously listed since 2002. The full data set covers the years 2002 through Q3-2013.
We are interested in-- among other things-- comparisons using benchmark neutral portfolios, and take SPX as the benchmark; our models are thus for the $m=401$-dimensional vector of returns comprising SPX as the first entry, followed by the 400 stocks. 
For each series $j$, daily log-returns are $y_{jt}= \log(\text{price}_{jt}/\text{price}_{j,t-1})$ using the daily closing prices.  
For clarity, as the SPX series is of particular interest as a benchmark, we label the first return series accordingly: $y_{\text{SPX},t}\equiv y_{1t}.$ 

Our comparative analyses assume that there are no bid-ask spreads, and that trading costs are in the amount of 10 basis points of the traded volume. We assume that all trades can be executed at the daily closing price and that short-selling is possible. Our calculations of annualized returns and volatilities assume that a year consists of 252 trading days.

\subsection{Forecast Model Specifications}

We study analysis of several variants of the SGDLM of Section~\ref{sgdlm2:sec:forecast-models}, based on different 
choices of exogenous predictors and discount factors. Table~\ref{sgdlm2:tb:data-example:DLMs} provides a full summary of the models used. Each SGDLM has fixed, core parental set sizes  $|pa(j)|=20$ for each series $j,$ and the adaptive parental strategy is based on $\Delta t=10.$

\begin{table}[htbp!]
\centering
\small
\begin{tabular}{c c c c c c c}
Model & Predictors & $\beta_j$ & $\delta_{j\phi}$ & $\delta_{j\gamma}$ & $\ell_{2003:2013}$ & $MAD_{2003:2013}$ \\ \hline  
M1 & \eqn{sgdlm2:eq:DLM-predictors-v0} & $0.95$ & $0.995$ & $0.995$ & $-661.6$ & $1.399 \times 10^{-2}$ \\
M2 & \eqn{sgdlm2:eq:DLM-predictors-v0} & $0.95$ & $0.995$ & $0.996$ & $-655.8$ & $1.401 \times 10^{-2}$\\
M3 & \eqn{sgdlm2:eq:DLM-predictors-v0} & $0.95$ & $0.995$ & $0.997$ & $-650.4$ & $1.408 \times 10^{-2}$ \\
M4 & \eqn{sgdlm2:eq:DLM-predictors-v0} & $0.95$ & $0.995$ & $0.998$ & $-646.1$ & $1.397 \times 10^{-2}$ \\
M5 & \eqn{sgdlm2:eq:DLM-predictors-v0} & $0.95$ & $0.995$ & $0.999$ & $-642.6$ & $1.398 \times 10^{-2}$ \\

MA1 & \eqn{sgdlm2:eq:DLM-predictors-v1} & $0.95$ & $0.995$ & $0.995$ & $-662.3$ & $1.402 \times 10^{-2}$ \\
MA2 & \eqn{sgdlm2:eq:DLM-predictors-v1} & $0.95$ & $0.995$ & $0.996$ & $-656.6$ & $1.404 \times 10^{-2}$ \\
MA3 & \eqn{sgdlm2:eq:DLM-predictors-v1} & $0.95$ & $0.995$ & $0.997$ & $-652.4$ & $1.491 \times 10^{-2}$ \\
MA4 & \eqn{sgdlm2:eq:DLM-predictors-v1} & $0.95$ & $0.995$ & $0.998$ & $-647.4$ & $1.402 \times 10^{-2}$ \\
MA5 & \eqn{sgdlm2:eq:DLM-predictors-v1} & $0.95$ & $0.995$ & $0.999$ & $-644.3$ & $1.410 \times 10^{-2}$ \\
%
\end{tabular}
\normalsize
\caption{List of SGDLMs  in case study: Predictors column indicates the model equation; $\beta_\ast,\delta_\ast$ are 
values of model discount factors; $l_\ast$ gives values of overall predictive log-likelihoods from the data analyses over 
2003--2013; $MAD_\ast$ gives corresponding mean absolute deviations of one-step ahead point forecast errors averaged over all 401 series
and across the period.} \label{sgdlm2:tb:data-example:DLMs}
\end{table}

\para{Dynamic linear model forms.} The simplest DLM form is the local-level model with
\begin{equation} \label{sgdlm2:eq:DLM-predictors-v0}
\bF_{jt} = \bF_t = 1
\end{equation}
for all $j=1{:}m$ and $t$. 
A first extension of the base model adds the average forecast error of the last 5 days as a predictor,
\begin{equation} \label{sgdlm2:eq:DLM-predictors-v1}
\bF_{jt} =  (1,x_{jt})' \quad \textrm{with}\quad  x_{jt} = 0.2 \sum_{k=1{:}5} (y_{j,t-k} - f_{j,t-k})
\end{equation}
where $f_{jt}$ is the one-step ahead point forecast for $y_{jt},$ namely the 
mean of the forecast distribution computed at time $t-1$. 
Note that this form already uses individual predictors for each series, which is not possible in the standard WDLM.

\begin{figure}[htbp!]
\centering
\includegraphics[width=6.5in]{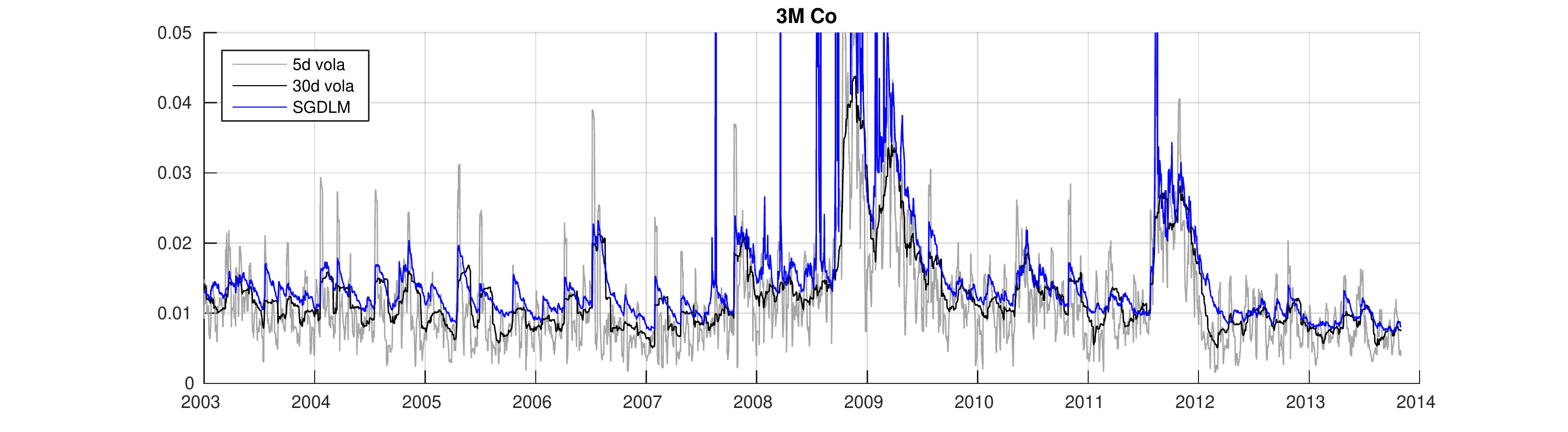}
\caption{Volatility of stock returns of company 3M $(j=245)$: observed 5-day and 30-day tracking volatilities (gray and black, respectively) together with the predicted volatility (blue) under model M1 of Table~\ref{sgdlm2:tb:data-example:DLMs}.} \label{sgdlm2:fig:volas-j245}
\centering
\includegraphics[width=6.5in]{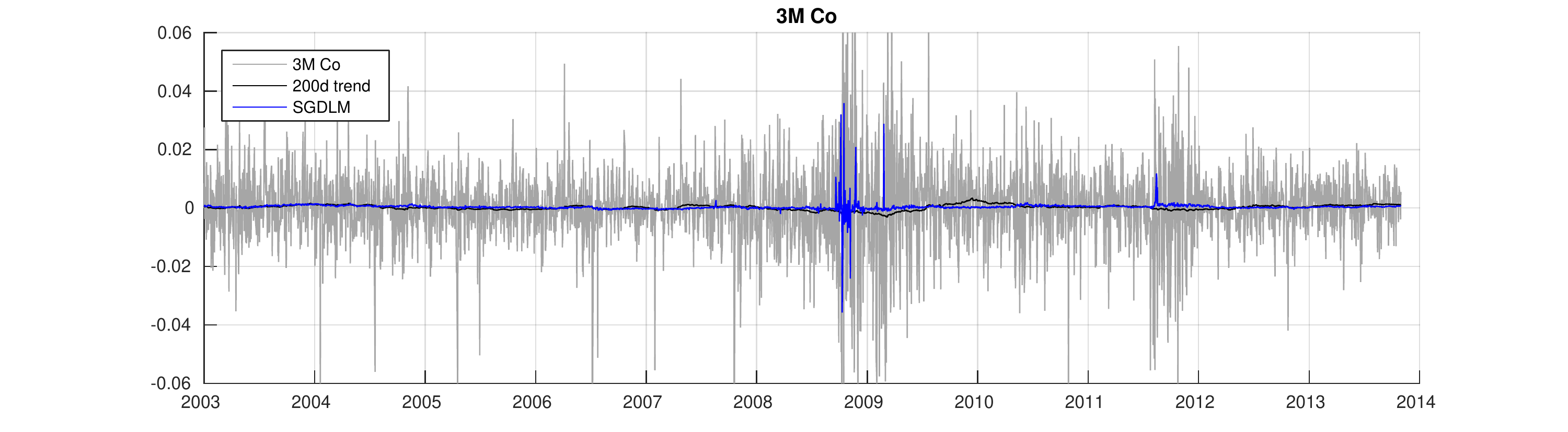}
\caption{Daily log-returns (gray) of company 3M, together with the observed 200-day trend line (black) and the predicted trend (blue) from model M1 of Table~\ref{sgdlm2:tb:data-example:DLMs}.} \label{sgdlm2:fig:means-j245}
\end{figure}

\para{Discount factors.} Values of discount factors close to 1 imply more stable trajectories of the stochastic variances (controlled by $\beta_j$) and dynamic state parameters (controlled by the
$\delta_\ast$).   Based on   past experience with earlier models (in foreign exchange
rates, stock and commodity studies; see earlier noted references)  
values of $\beta_j$ around $0.93-0.97$ are anticipated to be required to reflect residual volatilities, while
higher values of the $\delta_\ast$ parameters should be relevant in reflecting smaller stochastic changes in
the dynamic state parameters in our models with several parents for each series. 
Preliminary evaluation of predictive performance of our SGDLMs across a range of discount
values support this view, and a selection of summaries are reported here. Based on this preliminary
evaluation, we select $\beta_j = 0.95$ for all $j$ for the examples here. One aspect of this is that the resulting SGDLMs 
yield volatility predictions similar to the raw 30-day historical volatilities; see one example-- for the returns on 
stocks of company 3M-- in Figure~\ref{sgdlm2:fig:volas-j245}. The predicted volatility reacts instantaneously to market gyrations as can be seen by the volatility spikes on individual days during the most intense phase of the financial crisis. 

The discount factor for the dynamic regression coefficients $\bphi_{jt}$ is taken as $\delta_{j\phi} = 0.995$ for the current examples.   One aspect of this is that the resulting local trend forecasts are similar to the 200-day tracking moving average of returns, but show more responsiveness in times of more 
dramatic change during the financial crisis; see Figure~\ref{sgdlm2:fig:means-j245}.
Finally, discount factors $\delta_{j\gamma}$  for parental coefficients are assessed across values $\delta_{j\gamma} \in \{0.995, 0.996, 0.997, 0.998, 0.999\}$. In evaluating the discount factors, we balance a quantitative assessment of predictive abilities through predictive log-likelihoods and mean absolute deviation (MAD) with a qualitative assessment of desirable characteristics, and on how they impact on portfolio performance. 

\begin{figure}
\centering
\includegraphics[width=6in]{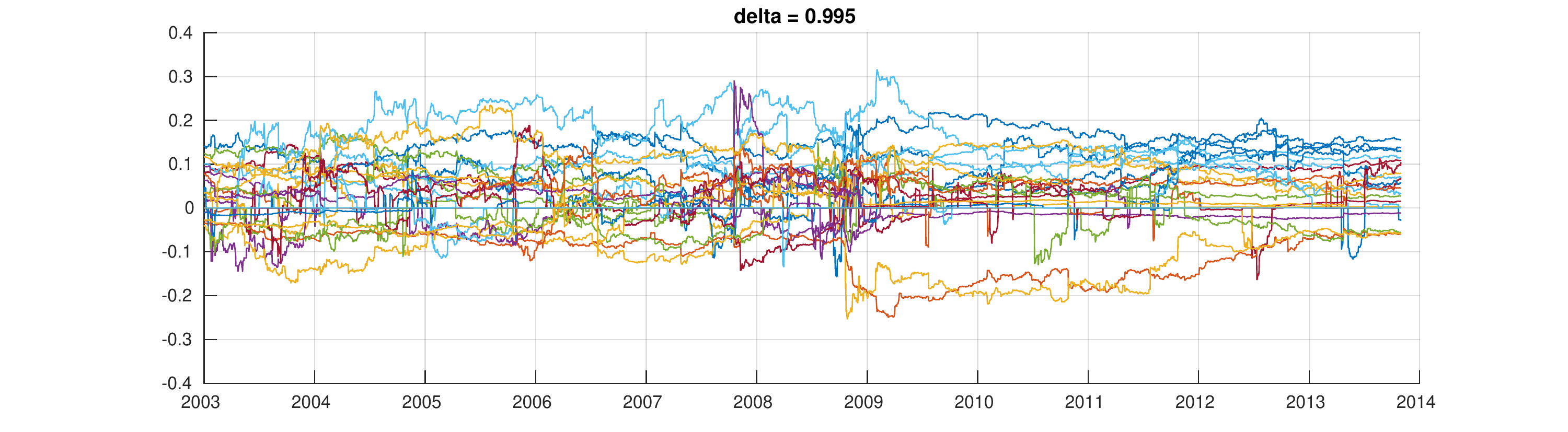} \\
\includegraphics[width=6in]{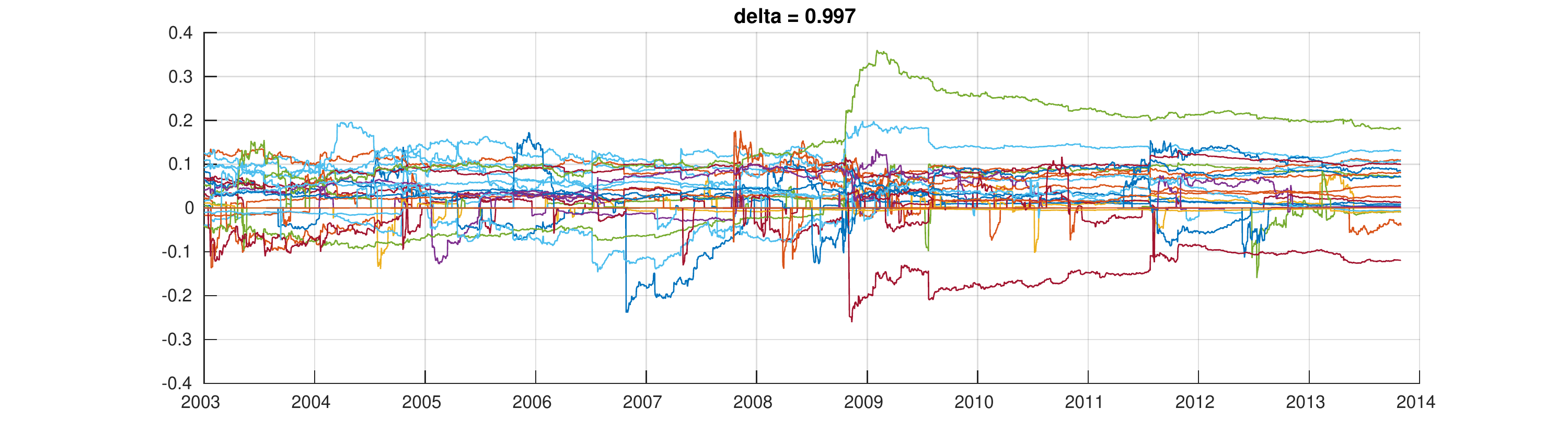} \\
\includegraphics[width=6in]{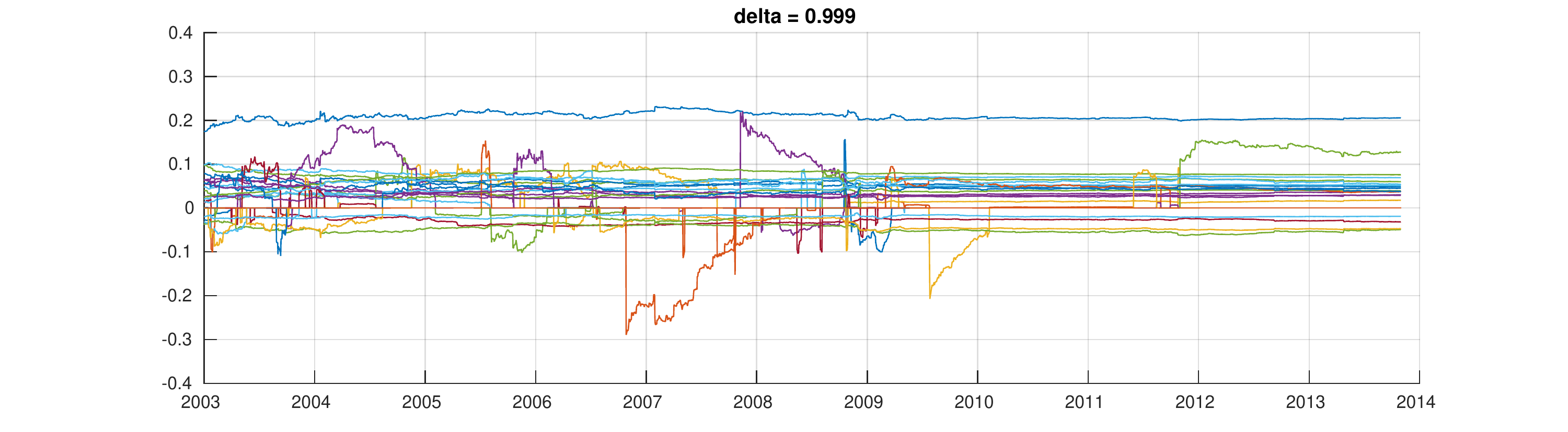} \\
\caption{Time trajectories of the on-line posterior means of dynamic parental regression coefficients
in models for 3M, using discount factors
$\delta_{j\gamma} \in \{0.995, 0.997, 0.999\}$ (models M1, M3, M5, respectively).} 
\label{sgdlm2:fig:MXVB_m-deltas}
\end{figure}

The predictive log-likelihoods (logs of model marginal likelihoods computed as the product of 1-step ahead
forecast densities over time)  increase with higher values of $\delta_{j\gamma}$, while raw point forecast accuracy as measured by MAD favours $\delta_{j\gamma}=0.997$ over higher and lower values; see Table~\ref{sgdlm2:tb:data-example:DLMs}. Visual inspection of the time evolution of the dynamic regression coefficients for the simultaneous parents shows that, as expected, higher discount values rapidly constrain
adaptivity in the parental coefficients and reduce the responsiveness of the model in times of more dramatic change; see  Figure~\ref{sgdlm2:fig:MXVB_m-deltas}. 
As a result, we recommend models with $\delta_{j\gamma}$ in the $0.995-0.997$ range 
for practical use. 

\subsection{Aspects of the Simultaneous Parental Structure}
Adaptive selection and evolution of parental sets 
of Section~\ref{sgdlm2:sec:sgdlm-bayes-hotspot} uses the $m=401$-dimensional local-level WDLM with 
state discount factor of 0.95 and multivariate volatility discount factor of  0.8. The latter induces a higher level responsiveness to changing covariance patterns, so engendering adaptability in the
generation of candidates for addition to  parental sets. 

\begin{figure}
\centering
\includegraphics[width=6in]{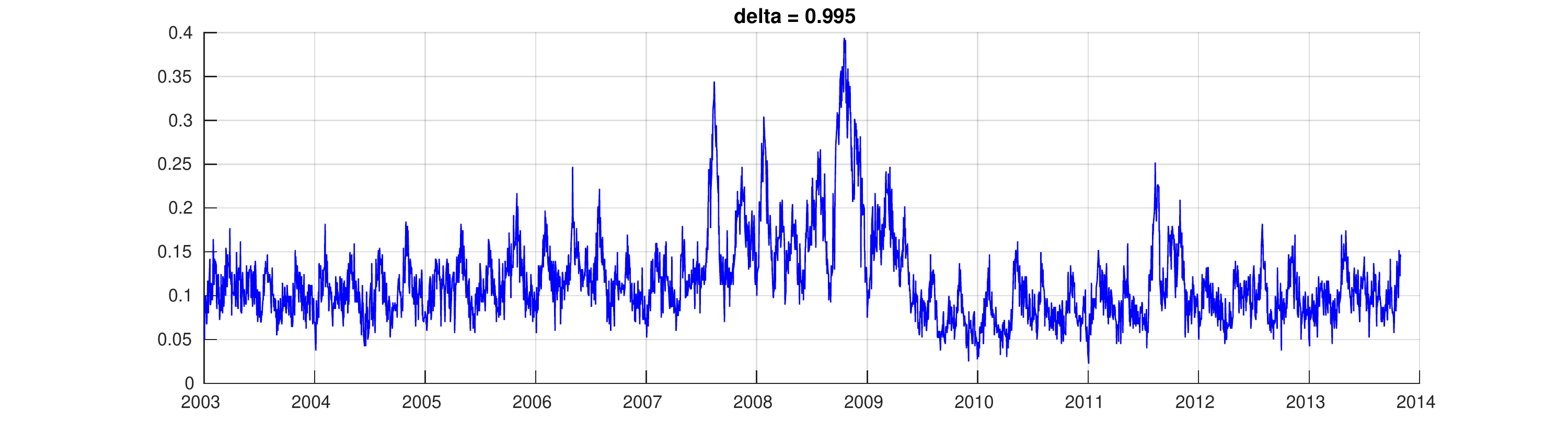} \\
\includegraphics[width=6in]{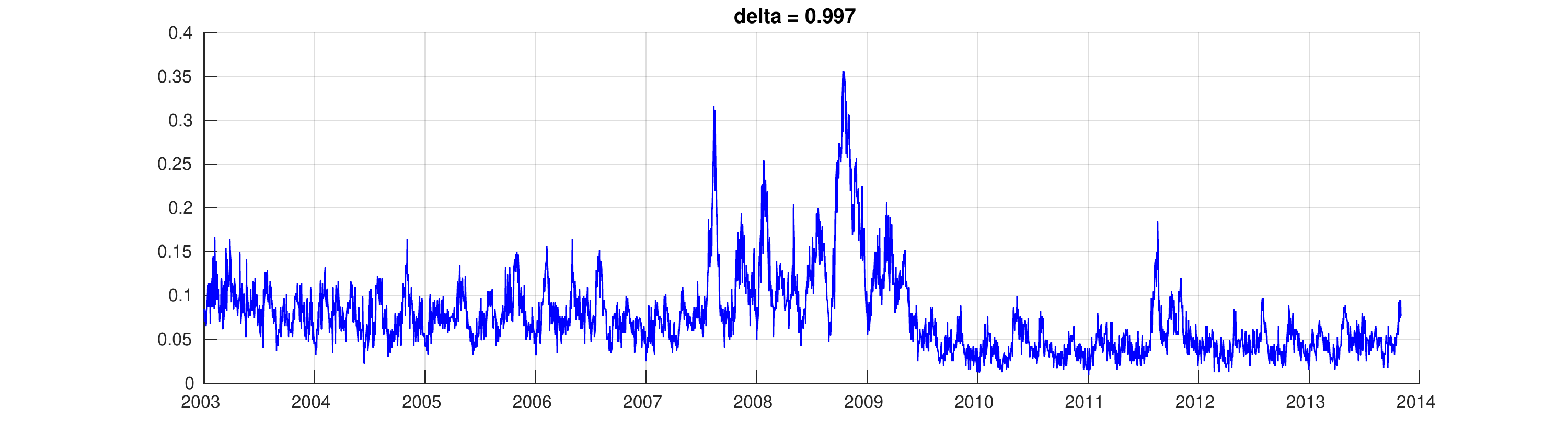} \\
\includegraphics[width=6in]{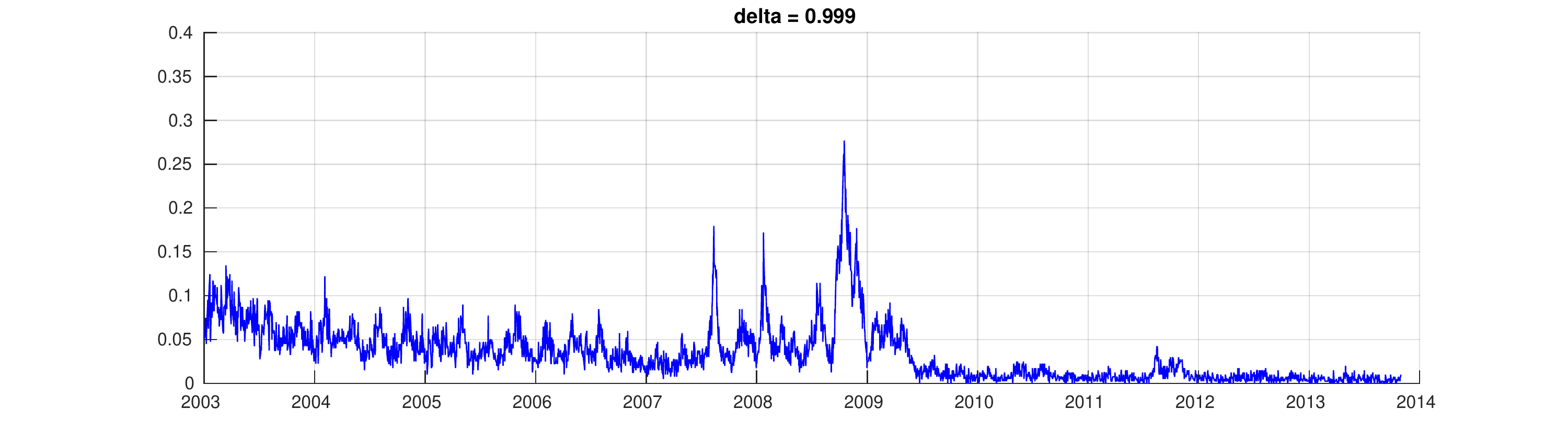} \\
\caption{Time trajectories of the  fractional changes in the core simultaneous parental sets across all $j=1{:}401$ series for discount factors 
$\delta_{j\gamma} \in \{0.995, 0.997, 0.999\}$ (models M1, M3, M5, respectively).   Thus, for example, a level of about 0.1 at any time $t$ indicates that about 40 of the 401 series had a change 
in core simultaneous parents at that time.
} 
\label{sgdlm2:fig:sp-changes-deltas}
\end{figure}

The SGDLM discount factor $\delta_{j\gamma}$  plays a key role in 
impacting the \lq\lq turnover rate" in the simultaneous parental sets using the adaptive strategy;
it directly influences the signal-to-noise ratio of dynamic regression coefficients of the incumbent simultaneous parents $sp_{\text{core},t}(j)$ and those of the proposed simultaneous parents in the warm-up set $sp_{\text{up},t}(j)$. Recall that a candidate simultaneous parent is accepted, if, after $\Delta t$ steps, its signal-to-noise ratio is greater than the smallest of any simultaneous parent in the core set $sp_{\text{core},t}(j)$: a smaller $\delta_{j\gamma}$ decreases the signal-to-noise ratios of parents in the core set.
Collective changes in the core sets of parents over time offer insight into model selection dynamics; see Figure~\ref{sgdlm2:fig:sp-changes-deltas}. This exemplifies the role of  $\delta_{j\gamma}$ that, when taking higher values, promotes lower levels of adaptation; see, in particular, the
lower frame of the figure over the  last few years, where the adaptation of parents decreases dramatically. 
To maintain responsiveness in this aspect of model specification, lower values of this
discount factor, in the range $0.995-0.997$ as in the upper two frames in the figure, 
are recommended.  Some more detailed visuals reflecting the time-evolution of the parental set 
of 3M 
appear in 
Figures~\ref{sgdlm2:fig:sp-j245-delta995-3MCo-all-sorted} and
\ref{sgdlm2:fig:sp-j245-delta995-3MCo-core-top-sorted} display 
of \ref{sgdlm2:app:sp-study}. 

\para{Cross-sector relationships.}  To further explore the behaviour of the dynamic parental set 
selection strategy in this study,  we consider the industrial sector relationships. 
Each of the S\&P stocks is classified into one of 9 industry sectors: basic materials, communications, consumer (cyclical), consumer (non-cyclical), energy, financial, industrial, technology, utilities. Table~\ref{sgdlm2:tb:sector-split} lists the distribution of the 400 stocks
across sectors.

We are now interested in structural links across industry sectors identified by our dynamic selection strategy. It appears intuitive to expect that the simultaneous parents $sp_t(j)$ of series $j$ will tend to be
enriched in stocks from the same sector as series $j$, while having  representatives of a few other sectors of systematic importance to  the main businesses of series $j.$ The fate of the energy sector, for example, 
naturally  depends on the well-being of the industrial sector that represents major energy consumers.  
Using model M1, we summarize such cross-sector connections in Table~\ref{sgdlm2:tb:sp-cross-sectors}; the values shown are deviations from a uniform distribution across all 400 candidate series.

\begin{table}[htp!]
\centering
\small
\begin{tabular}{l c c}
Sector & \# Member Series &  Frequency \\  \hline
Basic Materials & 20 & 5.00\% \\
Communications & 28 & 7.00\% \\
Consumer (cyclical) & 59 & 14.75\% \\
Consumer (non-cyclical) & 73 & 18.25\% \\
Energy & 41 & 10.25\% \\
Financial & 67 & 16.75\% \\
Industrial & 56 & 14.00\% \\
Technology & 35 & 8.75\% \\
Utilities & 21 & 5.25\% \\
\end{tabular}
\normalsize
\caption{Distribution of the 400 S\&P member stocks across industry sectors.} \label{sgdlm2:tb:sector-split}
\bigskip\bigskip
\centering
\small
\begin{tabular}{l r r r r r r r r r r}
Sector of $j$ & \rot{Basic Materials} & \rot{Communications} & \rot{\parbox{1in}{Consumer\\(cyclical)}} & \rot{\parbox{1in}{Consumer\\(non-cyclical)}} & \rot{Energy} & \rot{Financial} & \rot{Industrial} & \rot{Technology} & \rot{Utilities} & \rot{S\&P 500} \\  \hline
B. Materials &\CG $\mathbf{+62\%}$ &\CR $-21\%$ &\CR $-19\%$ &\CR $-9\%$ & $+2\%$ & $+1\%$ & $-5\%$ &\CG $+15\%$ &\CG $+39\%$ &\CG $+55\%$ \\
Comms. &\CR $-7\%$ & \CR $\mathbf{-23\%}$ &\CG $+18\%$ & $-1\%$ & $+4\%$ & $0\%$ & $+1\%$ &\CR $-19\%$ & $+4\%$ &\CG $+118\%$ \\
Cyclicals &\CR $-13\%$ & $+4\%$ & $\mathbf{+5\%}$ & $-5\%$ &\CR $-9\%$ & $+1\%$ &\CG $+15\%$ &\CR $-13\%$ & $+1\%$ &\CG $+104\%$ \\
Non-cyclicals & $+3\%$ &\CG $+9\%$ & $-3\%$ & $\mathbf{-2\%}$ & $-3\%$ & $+3\%$ & $+4\%$ &\CR $-17\%$ & $+3\%$ &\CG $+220\%$ \\
Energy &\CG $+13\%$ &\CG $+7\%$ &\CG $+9\%$ & $+1\%$ & $\mathbf{+2\%}$ &\CR $-10\%$ &\CR $-17\%$ & $+2\%$ &\CG $+16\%$ &\CG $+138\%$ \\
Financial &\CG $+12\%$ & $-5\%$ & $-3\%$ &\CG $+7\%$ & $-4\%$ & $\mathbf{+5\%}$ & $-2\%$ &\CR $-8\%$ &\CR $-14\%$ &\CG $+137\%$ \\ 
Industrial & $-3\%$ &\CR $-10\%$ &\CG $+11\%$ &\CG $+7\%$ &\CR $-13\%$ & $+2\%$ & $\mathbf{-5\%}$ & $-1\%$ &\CR $-9\%$ &\CG $+111\%$ \\
Technology &\CG $+9\%$ &\CR $-10\%$ &\CR $-7\%$ & $-4\%$ &\CG $+16\%$ & $+5\%$ & $-4\%$ &\CR $\mathbf{-13\%}$ &\CG $+14\%$ &\CG $+210\%$ \\
Utilities &\CG $+22\%$ &\CR $-23\%$ & $-4\%$ & $-3\%$ &\CG $+12\%$ &\CR $-13\%$ & $+2\%$ &\CR $-10\%$ &\CG $\mathbf{+54\%}$ &\CG $+200\%$ \\
S\&P 500 &\CR $-100\%$ &\CG $+84\%$ &\CR $-6\%$ &\CG $+25\%$ &\CG $+41\%$ &\CR $-23\%$ &\CR $-79\%$ &\CG $+28\%$ &\CR $-75\%$ & \multicolumn{1}{c}{\textbf{n/a}} \\
\end{tabular}
\normalsize
\caption{Relative deviations of industry clusters indicated by the selected simultaneous parents in the core sets from a uniform distribution over all 400 candidate series. Results are from model M1 of Table~\ref{sgdlm2:tb:data-example:DLMs}.} \label{sgdlm2:tb:sp-cross-sectors}
\end{table}

We see that  increased simultaneous parental links between stocks from the same industry sector appear only in the basic materials and utilities sectors;  substantially smaller  within-sector 
representations arise  in the communications and technology sectors. Generally, these examples define two dominating sectors as well as two under-represented sectors: basic materials and utilities are much more strongly represented across all series, while communications and technology sectors are significantly under represented.  Of other sectors, non-cyclicals, industrials and financials are slightly under-represented, and the rest are mixed. The strong deviation from a uniform distribution of parents across sectors provides reason to conclude that the observed sectorial clusters reflect structural links between the corresponding sectors in the real economy. That said, the single most heavily over-represented series of all $m=401$ is the S\&P 500, which is rather unsurprising, as all series are member stocks of that index and underlie some shared market dynamics.

\subsection{Entropy as a Measure of Market Stress}
The St.~Louis Federal Reserve Bank Financial Stress Index (STLFSI) was created in early 2010 to measure financial stress in the market by a weighted average of 18 weekly data series; the weights of each index are determined by principal components analysis; see the appendix in~\citet{kliesensmith2010}. Seven of these series are on interest rates, six are on yield spreads, and five are on other indicators. The index is designed to have an average value of 0, with positive readings indicating above-average stress, and negative readings indicating below-average stress. A new index reading is published each week~\citep{stlouisfed-stlfsi}.

\begin{figure}[hbpt!]
\centering
\includegraphics[width=6in]{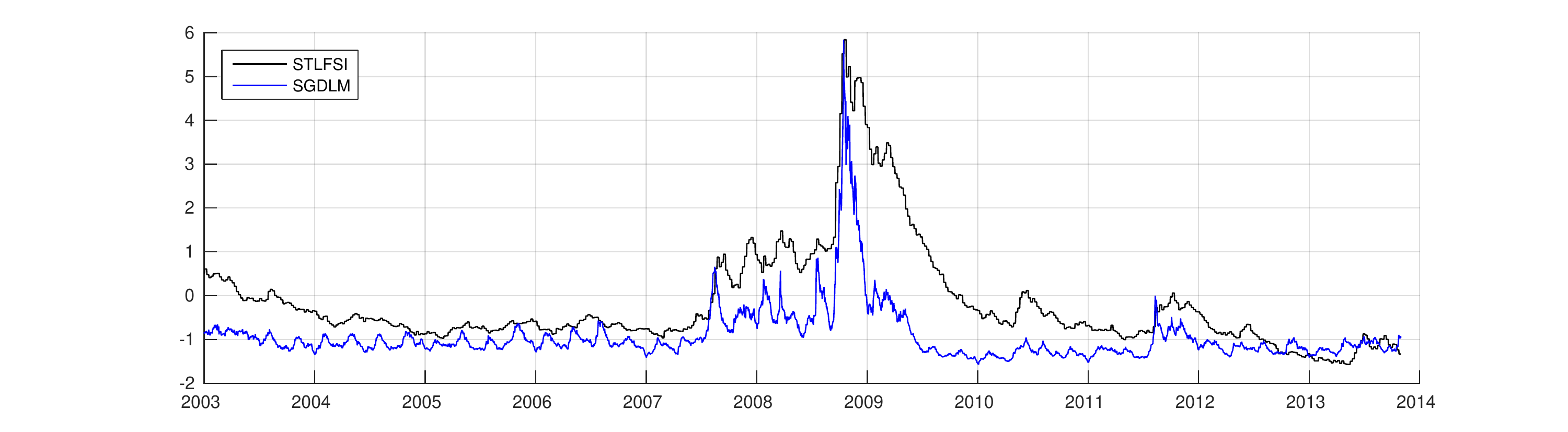}
\caption{STLFSI index (black), and scaled and translated entropy under model M1 (blue).} \label{sgdlm2:fig:entropy-stlfsi}
\end{figure}

We have found that one purely statistical measure arising from our SGDLM analysis 
of the S\&P data has a surprisingly strong relationship to the STLFSI.  Our measure is 
simply the optimized entropy measure emerging from our importance sampling/variational Bayes (VB)
analysis-- i.e., at each time $t$, 
the minimum value of the KL divergence in the decouple step of the SGDLM analysis. 
Scaling and inverting this direct entropy measure to map to the STLFSI scale, we know that the transformed measure will be low when the posterior VB approximation is very accurate, and will increase when the approximation
breaks down. The overlaid plots in Figure~\ref{sgdlm2:fig:entropy-stlfsi}
 show a strong concordance with the econometrically
derived STLFSI.   We therefore suggest a purely observational interpretation of 
 our entropy metric as a measure of stock market stress.  We  observe a few interesting features in the
figure.  First, after the peak in fall 2008, stress in the stock market fell substantially more rapidly than stress in the broader financial markets. Second, overall financial market stress levels diverged from stock market stress levels around June 2010 and through the Eurozone crisis period.  After that until the end of 2014, the 
two stress areas come together at lower, more stable levels.  

The relationship between our SGDLM entropy and the STLFSI financial is remarkable. Our measure is based solely on statistical analysis of the set of S\&P  series; the S\&P 500 index is only one of 18 series affecting the STLFSI, and the majority of the other indices used are not directly linked to the stock market. \cite{kliesensmith2010} find that weekly updates of the STLFSI provide improvements over the 
monthly updates of another traditional stress index, that of the Kansas City Federal Reserve Bank. 
One rationale is the high pace at which significant developments in financial markets impact the global economy. We therefore suggest that a focus on 
daily updates can be expected to improve information flows further, and note that our 
entropy metric is updated on a daily basis in this case study.  Further inspection shows that it does in fact lead 
both the Kansas City Fed  and St. Louis Fed indices. 
Thus, in addition to defining a key  monitor on the SGDLM model adequacy-- with an ability to signal the need for potential interventions 
at times of increased change-- the new entropy metric has the clear potential to add to the understanding of 
market dynamics in terms of global risk measures, providing daily updates that may be of
use and importance to all market participants-- consumers, investors, and regulators alike. 

\subsection{Portfolio Comparisons}
Using the portfolios defined in Section~\ref{sgdlm2:sec:investment-strategy}, we
analyze the optimization-based investment strategies summarized in Table~\ref{sgdlm2:tb:data-example:investment-rules}. This pairs each of the 10 SGDLMs  in Table~\ref{sgdlm2:tb:data-example:DLMs} with each of the 6 quantitative investment strategies. For baselines, we include the S\&P 500 index itself, along with a simple equally-weighted portfolio. We conduct our investment study as follows: (i) at market close on day $t-1$, update the model distributions based on the observation $\by_{t-1}$; (ii) compute or simulate the 1-step ahead forecast distribution for $\by_t$; solve the portfolio optimization and adjust the portfolio investment weight vector $\bw_t$ to the new, optimized value; (iii) move to time $t$, observe and record the realized returns, and continue with $t \to t+1$.  In addition to exploring the set of SGDLMs, we add comparison 
with the standard WDLM using a local level forecasting component for each series.  A set of 5 WDLMs, labeled W1-W5, differ only in the values of the discount factors used for this standard benchmark model. Each uses a Wishart volatility matrix discount factor of 0.95,  while they differ in the local local discount factor which takes 
values 0.995,0.996,0.997,0.998,0.999 in moving from W1 to W5, respectively. 
\begin{table}[htbp!]
\centering
\small
\begin{tabular}{ll}
Strategy& Description \\  \hline
\,\,\, SPX & passive investment in the S\&P 500 \\
\,\,\, P0 & equal weights \\
\,\,\, P1* & minimum variance \\
\,\,\, P2* & target return $\tau_t = 10\%/252$ \\
\,\,\, P3* & target return $\tau_t = 15\%/252$ \\
\,\,\, P4* & SPX neutral, minimum variance \\
\,\,\, P5* & SPX neutral, target return $\tau_t = 10\%/252$ \\
\,\,\, P6* & SPX neutral, target return $\tau_t = 15\%/252$ \\
\end{tabular}
\normalsize
\caption{Portfolio investment strategies compared in S\&P study.} \label{sgdlm2:tb:data-example:investment-rules}
\end{table}

Our investment rules P1*-P6* extend the pure minimum variance rules (potentially including target return and benchmark-neutral constraints) with a dynamic churn reduction mechanism: whenever the expected gain of updating the portfolio weights $\bw_t$ to the time $t+1$ weights indicated by the optimization rule, $\bw_{t+1}^0$, does not outweigh the trading cost of that update, the portfolio weights are only updated to the extent that the expected gain from updating to $\bw_{t+1} = \lambda \bw_t + (1-\lambda) \bw_{t+1}^0$, $\mu_{t+1}' (\bw_{t+1} - \bw_t)$ equals the trading cost $c |\bw_{t+1} - \bw_t|$.

\begin{table}[htbp!]
\centering
\small
\begin{tabular}{c c c c c c c c c}
Model & \blue SPX & \red P0 & P1* & P2* & P3* & P4* & P5* & P6* \\  \hline
     & \blue 0.31 & \red 0.41  \\
M1 & & & 0.71 & 0.79 & 0.86 & 0.71 & 0.76 & 0.84 \\ 
M2 & & & 0.82 & 0.87 & 0.86 & 0.81 & 0.87 & 0.88 \\
M3 & & & 0.63 & 0.62 & 0.69 & 0.60 & 0.62 & 0.69 \\
M4 & & & 0.68 & 0.68 & 0.67 & 0.70 & 0.69 & 0.69 \\
M5 & & & 0.73 & 0.74 & 0.72 & 0.68 & 0.68 & 0.67 \\
MA1& & & 0.72 & 0.71 & 0.69 & 0.71 & 0.74 & 0.73 \\
MA2& & & 0.77 & 0.74 & 0.70 & 0.76 & 0.72 & 0.68 \\
MA3& & & 0.80 & 0.76 & 0.68 & 0.78 & 0.72 & 0.68 \\
MA4& & & 0.83 & 0.78 & 0.75 & 0.83 & 0.78 & 0.75 \\
MA5& & & 0.88 & 0.86 & 0.69 & 0.91 & 0.88 & 0.77 \\
W1 & & & 0.06 &-0.01 &-0.11 & 0.10 & 0.01 &-0.09 \\
W2 & & & 0.06 &-0.00 &-0.10 & 0.09 & 0.02 &-0.08 \\
W3 & & & 0.05 & 0.01 &-0.08 & 0.09 & 0.03 &-0.06 \\
W4 & & & 0.05 & 0.04 &-0.03 & 0.08 & 0.05 &-0.01 \\
W5 & & & 0.05 & 0.06 & 0.03 & 0.08 & 0.08 & 0.05 \\
\end{tabular}
\normalsize
\caption{Sharpe ratios of portfolios SPX, P0, P1*-P6* based on forecasts from models M1-M5, MA1-MA5, and W1-W5 from 2003--2013. }   \label{sgdlm2:tb:data-example:investment-sharpe-all}
\end{table}

The annualized portfolio Sharpe ratios of the investment strategies in Table~\ref{sgdlm2:tb:data-example:investment-rules} are summarized in Table~\ref{sgdlm2:tb:data-example:investment-sharpe-all}; the portfolio returns and volatilities are listed in Tables~\ref{sgdlm2:tb:data-example:investment-returns-all} and~\ref{sgdlm2:tb:data-example:investment-volas-all} of \ref{sgdlm2:app:sp-study}. All portfolios and models provide a better risk{:}return profile than a passive investment in the S\&P 500 index. In fact, all dynamic minimum variance portfolios (P1*-P6*) yield at least twice the return for every unit volatility than the S\&P 500. The local level models M1-M5 perform best with lower discount factors $\delta_{j\gamma} \in \{0.995, 0.996\}$; furthermore, model M1 is the only one whose portfolios show better return vs. volatility characteristics as the target returns are raised! Based on models MA1-MA5, the minimum variance portfolios without a target return perform better for higher discount factors $\delta_{j\gamma}$; this trend weakens as increasingly ambitious target returns are introduced, which then lead to declining portfolio performance due to 
increased risk.

An interesting finding is that the best-performing models in terms of predictive log-likelihoods or mean absolute deviations (Table~\ref{sgdlm2:tb:data-example:DLMs}) are not the best models to use for investment decisions. This latter point is consistent with experience in other areas of statistical and decision analytic work, where utility-guided selection of models can lead to different model structures than those favored on purely statistical metrics~\citep[e.g.][]{JonesDobraCarvalhoHansCarterWest2005, CarvalhoWest2007}.

\begin{figure*}[htbp!]
\begin{center}
\includegraphics[width=6in]{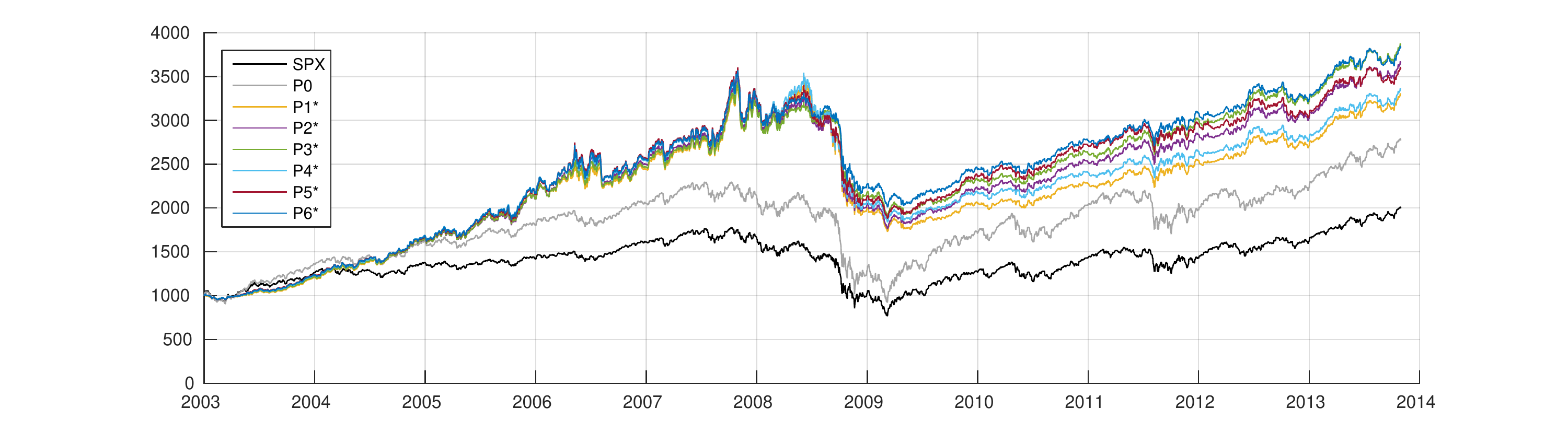}
\caption{Trajectory of portfolio value over 2003--2013 based on forecasts from model M1.} \label{sgdlm2:fig:data-example:pf-process}
\end{center}
\begin{center}
\includegraphics[width=6in]{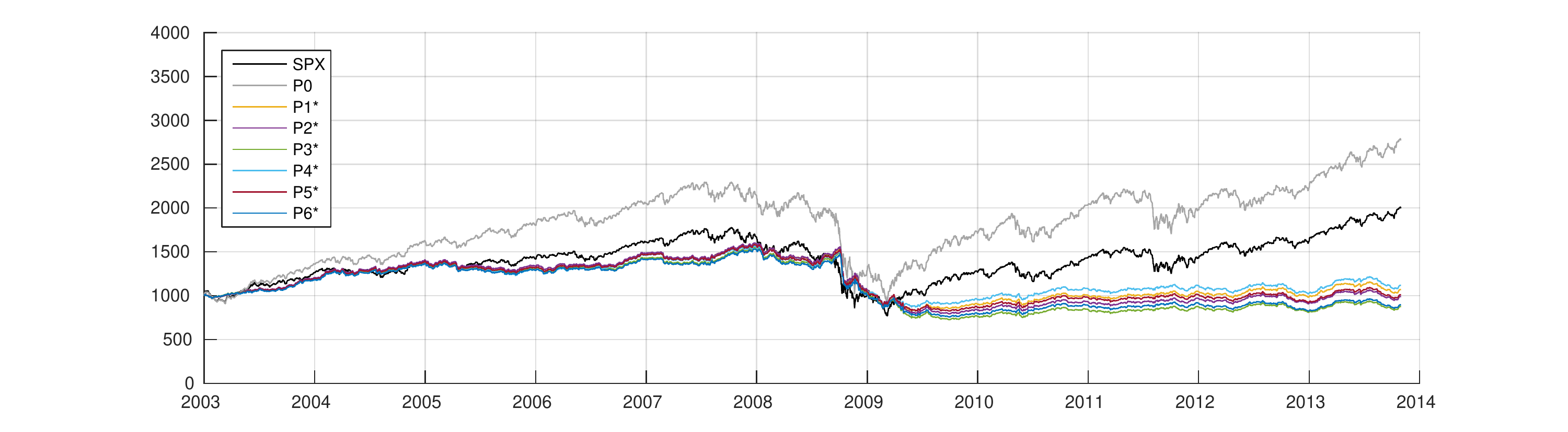}
\caption{Trajectory of portfolio value over 2003--2013 based on forecasts from model W1.} \label{sgdlm2:fig:data-example:pf-process-VD}
\end{center}
\end{figure*}

Figure~\ref{sgdlm2:fig:data-example:pf-process} graphs trajectories over time of the 
portfolio value for all portfolio strategies driven by model M1. Figure~\ref{sgdlm2:fig:data-example:pf-process-VD} shows the same for the corresponding WDLM W1. Under strategy P3*, \$1,000 invested at the beginning of 2003 would have grown to \$3,862 after accounting for 10bp trading costs; the range of the final value of investment P3* when based on models M2-M5 is from \$2,719 to \$3,545. The addition of the SPX benchmark neutrality constraint comes at the cost of a small decrease of performance, with portfolio values of P6* ranging from \$2,721 to \$3,830 based on models M1-M5. None of our portfolios generated annualized portfolio returns over 1\% when the forecasts from our WDLMs W1-W5 were used to derive investment decisions. The  
best-performing combination of WDLM and portfolio rule, W1 and P4*, would have grown \$1,000 into \$1,168 after accounting for the same 10bp trading costs during our investment horizon from 2003 to 2013. To put these numbers in perspective: a passive investment in the S\&P 500 would have grown into \$1,996 during the same time period. This example shows that the SGDLM vastly out-performs the standard WDLM as a model for investment decisions, and that its adoption can lead to significant monetary gains compared to model-free benchmarks such as an equal weights or passive index investment strategies as well as over WDLM-based investment decisions.

\section{Summary Comments} \label{sgdlm2:sec:final-remarks}
Our S\&P study investigated the multivariate forecast performance of the SGDLM for use in short-term forecasting and Bayesian portfolio decision analysis. The sparse, dynamic graphical model structure induced by the dynamic simultaneous parental predictor construct defines a parsimonious and potentially effective approach to structuring the contemporaneous relations in dynamic models. That is, the number of time-varying parameters to describe the structure of multivariate volatility is substantially reduced relative to standard models, which include the class of WDLMs. Data-respected and informed sparsity patterns, and adaptivity in representing such patterns as they may change over time, has the potential to improve forecasting accuracy and decisions based on such forecasts. The S\&P study results bear out this potential. The SGDLM modeling approach delivers substantially improved characterizations of volatility and co-volatility, in terms of forecast accuracy as well as usability in decision processes. The latter point is clearly highlighted in our portfolio investment evaluations. Portfolios reliant on model-based forecast information yield consistently higher nominal and risk-adjusted returns relative to standard approaches, and desired optimization constraints are more reliably achieved. As just one take-home summary to add to the more detailed results discussed in the study above, we note that the empirical performance reported here shows average annualized investment returns of SGDLM-driven quantitative investment rules as high as 12.1\% over a eleven year period from January 2003 to September 2013; that is a period during which the annual gains of the S\&P 500 averaged only 6.2\% (log-returns normalized to a 252 day year).  In parallel,  the framework as applied to the set of stock market 
series has generated a remarkably interesting metric of stock market risk that, computed as daily
markets close, has  potential as a leading indicator risk metric for use by financial 
and economic communities.

\appendix

\section{Wishart DLM (WDLM)} \label{sgdlm2:subsec:wdlm}
The WDLM  variant of the traditional Beta-Bartlett Wishart discounting model~\citep{QuintanaLourdesAguilarLiu2003} is the canonical multivariate extension of the univariate DLM that uses a common set of predictors across all series, multivariate normal observation errors, and replaces the univariate model's conjugate normal/Gamma priors for the states $\btheta_t$ and precisions $\lambda_t$ with conjugate matrix normal/inverse Wishart priors for the states $\bTheta_t$ and covariances $\bSigma_t$. Full details are in~\citet[][sect. 10.4.8]{PradoWest2010} whose notation we adopt here.

\para{Model specification.} The $m$-dimensional time series $\by_t := (y_{1t}, \ldots, y_{mt})'$ is modelled via
\begin{align}
\by_t' &= \bF_t' \bTheta_t + \bnu_t' \text{,}  &\bnu_t &\sim N(\bzero, \bSigma_t) \text{,} \label{sgdlm2:eq:WDLM-obs-eq} \\
\bTheta_t &= \bG_t \bTheta_{t-1} + \bOmega_t \text{,} &\bOmega_t &\sim N(\bzero, \bW_t, \bSigma_t) \text{,} \label{sgdlm2:eq:WDLM-evo-eq}
\end{align}
where: $\bF_t$ is a $p$-dimensional predictor vector; the $p \times m$ state matrix $\bTheta_t$ evolves according to \eqn{sgdlm2:eq:WDLM-evo-eq}; the $m \times m$ matrix $\bSigma_t$ is a time-varying volatility matrix; $\bOmega_t$ is a matrix normal innovation; $\bG_t$ is a known $p\times p$ state transition matrix; and $\bW_t$ is a known $p \times p$ innovation variance matrix. Each univariate element $y_{jt}$ of $\by_t$ then follows the model
\begin{align}
y_{jt} &= \bF_t'\btheta_{jt} + \nu_{jt} \text{,} &\nu_{jt} &\sim N(0, \sigma_{jjt}) \text{,} \\
\btheta_{jt} &= \bG_t \btheta_{j,t-1} + \bomega_{jt} \text{,} &\bomega_{jt} &\sim N(\bzero, \sigma_{jjt} \bW_t) \text{,}
\end{align}
where, for each $j=1{:}m$: the state evolution vectors $\bomega_{jt}$ are the columns of $\bOmega_t$; the state vectors $\btheta_{jt}$ are the columns of $\bTheta_t;$ and $\sigma_{jjt}$ is the $j-th$ diagonal element of $\bSigma_t.$ Non-zero covariances in $\bSigma_t$ induce cross-series dependencies via $C(\nu_{it}, \nu_{jt}) = \sigma_{ijt}$ and $C(\bomega_{it}, \bomega_{jt}) = \sigma_{ijt} \bW_t$ for $i \neq j$.

Key analysis components involve one-step evolution, forecasting and updating, as follows. For each time $t$, denote by $\cD_t$ all available information, here assumed to comprise just the past data with $\cD_t = \{ \cD_{t-1},\by_t\}$. 

\para{Prior at time $t$.} The prior for $\bTheta_t, \bSigma_t$ given $\cD_{t-1}$ is
\begin{equation} \label{sgdlm2:eq:WDLM-prior}
(\bTheta_t, \bSigma_t | \cD_{t-1}) \sim NIW(\ba_t, \bR_t, r_t, \bB_t),
\end{equation}
a matrix normal, inverse Wishart distribution. Here $\ba_t$ is the $p\times m$ prior mode of $\bTheta_t$ and $\bR_t$ the $p\times p$ within-column covariance matrix; the conditional prior of $\bTheta_t|\bSigma_t$ is matrix normal $N(\ba_t,\bR_t,\bSigma_t).$ Parameter $r_t > 0$ is the prior degrees-of-freedom, and $\bB_t$ is the $m\times m$ prior sum-of-squares matrix of the marginal inverse Wishart prior $\bSigma_t \sim IW(r_t,\bB_t)$; the prior mean of $\bSigma_t$ is $\bB_t/(r_t-2)$. 

\para{Forecasts at $t$.} Integration of~\eqn{sgdlm2:eq:WDLM-obs-eq} with respect to $p(\bTheta_t,\bSigma_t|\cD_{t-1})$ in~\eqn{sgdlm2:eq:WDLM-prior} yields the multivariate T, one-step forecast distribution $\by_t | \cD_{t-1} \sim T_{r_t}(\boldf_t, \bQ_t)$ with $r_t$ degrees of freedom, mode $\boldf_t = \bF_t' \ba_t$ and scale matrix $\bQ_t = q_t \bB_t / r_t$ where $q_t =1+ \bF_t' \bR_t \bF_t$; the forecast variance matrix is $q_t \bB_t / (r_t-2)$ if $r_t>2$. This is trivially extended to $k-$step ahead predictions. 

\para{Posterior at $t$.} The posterior of $\bTheta_t, \bSigma_t$ follows conjugate analysis upon observation of $\by_t$,
\begin{equation} \label{sgdlm2:eq:WDLM-posterior}
(\bTheta_t, \bSigma_t | \cD_t) \sim NIW(\bm_t, \bC_t, n_t, \bD_t) \text{,}
\end{equation}
with updated parameters $\bm_t = \ba_t + \bA_t \be_t$, $\bC_t = \bR_t - \bA_t \bA_t' q_t$, $n_t = r_t + 1$, and $\bD_t = \bB_t + \be_t \be_t' / q_t$ based on adaptive coefficient vector $\bA_t = \bR_t \bF_t / q_t$ and forecast error vector $\be_t = \by_t - \boldf_t$.

\para{Evolution to time $t+1$.} In moving ahead to time $t+1$, the posterior~\eqn{sgdlm2:eq:WDLM-posterior} evolves to the implied prior of the form of \eqn{sgdlm2:eq:WDLM-prior} but with index $t\to t+1$. For the DLM state matrix $\bTheta_{t+1}$, this involves the parameters $\ba_{t+1} = \bG_{t+1} \bm_t$ and $\bR_{t+1} = \bG_{t+1} \bC_t \bG_{t+1}' + \bW_{t+1}$. Here we specify the innovation variance matrices $\bW_{t+1} := \bG_{t+1} \bC_t \bG_{t+1}' (1/\delta-1)$ based on a single discount factor $\delta \in (0,1)$, so that $\bR_{t+1} = \bG_{t+1} \bC_t \bG_{t+1}'/\delta$. All models used in the case study below (Section~\ref{sgdlm2:sec:data-example}) are based on multivariate random walk evolutions for the states: $\bG_{t+1}=\bI$, so $\ba_{t+1} = \bm_t$ and $\bR_{t+1} = \bC_t /\delta$; that is, the states vary stochastically over time, but the evolution model does not anticipate directional variation. For the volatility matrix $\bSigma_{t+1}$, the variant of the Beta-Bartlett Wishart volatility model~\citep{PradoWest2010} implies trivially evolved parameters $r_{t+1} = \beta n_t$ and $\bB_{t+1} = \bD_t(r_{t+1} + m - 1) / (n_t + m - 1)$ akin to a random walk evolution of volatilities and co-volatilities. Here $\beta \in (0,1)$ is the discount factor governing the extent of stochastic changes in the evolution $\bSigma_t \to \bSigma_{t+1}$.

\section{Posterior Parameters from VB Decoupling} \label{sgdlm2:app:vb-decoupling}

In the SGDLM decoupling for evolution in Section~\ref{sgdlm2:subsec:sgdlm}, the parameters of the $m$ decoupled normal/gamma posteriors of~\eqn{sgdlm2:eq:VBapproxpostt} are computed to minimize the Kullback-Leibler divergence of the decoupled product of these normal/gammas from the full joint posterior, based on the importance sample generated from the latter. This follows~\citet[][sect. 12.3]{WestHarrison1997} and is an example of recently popularized---and more general---variational Bayes (VB) strategies~\citep[e.g.][]{JaakolaJordan2000, WandOrmerodPadoanFurhwirth2011}. 
 
The resulting quantities $(\bm_{jt}, \bC_{jt}, n_{jt}, s_{jt})$ are, for each $j=1{:}m,$ given as follows \citep[Section~3.2]{gruberwest2015a}:
\begin{itemize}
\item $\bm_{jt} = E[\lambda_{jt} \btheta_{jt}] / E[\lambda_{jt}]$;
\item $\bV_{jt} = E[\lambda_{jt} (\btheta_{jt}-\bm_{jt})(\btheta_{jt}-\bm_{jt})']$;
\item $d_{jt} = E[\lambda_{jt} (\btheta_{jt} - \bm_{jt})' \bV_{jt}^{-1} (\btheta_{jt} - \bm_{jt})]$;
\item $n_{jt}$ is is trivially calculated numerically as the unique value that satisfies 
 \newline\phantom{.}\,\,\,\, $\log(n_{jt} + p_j - d_{jt}) - \psi(n_{jt}/2) - (p_j - d_{jt}) / n_{jt} - \log( 2 E[\lambda_{jt}]) + E[\log \lambda_{jt}] = 0$; 
\item $s_{jt} = (n_{jt} + p_j - d_{jt}) / (n_{jt} E[\lambda_{jt}]);$ 
\item $\bC_{jt} = s_{jt} \bV_{jt}$.
\end{itemize}
The expectations here are evaluated by Monte Carlo based on the full posterior importance sample generated at the previous posterior recoupling/importance sampling step. 

\newpage

\section{S\&P 500 Case Study: Additional Analyses} \label{sgdlm2:app:sp-study}

\begin{figure}[htbp!]
\centering
\includegraphics[width=6in]{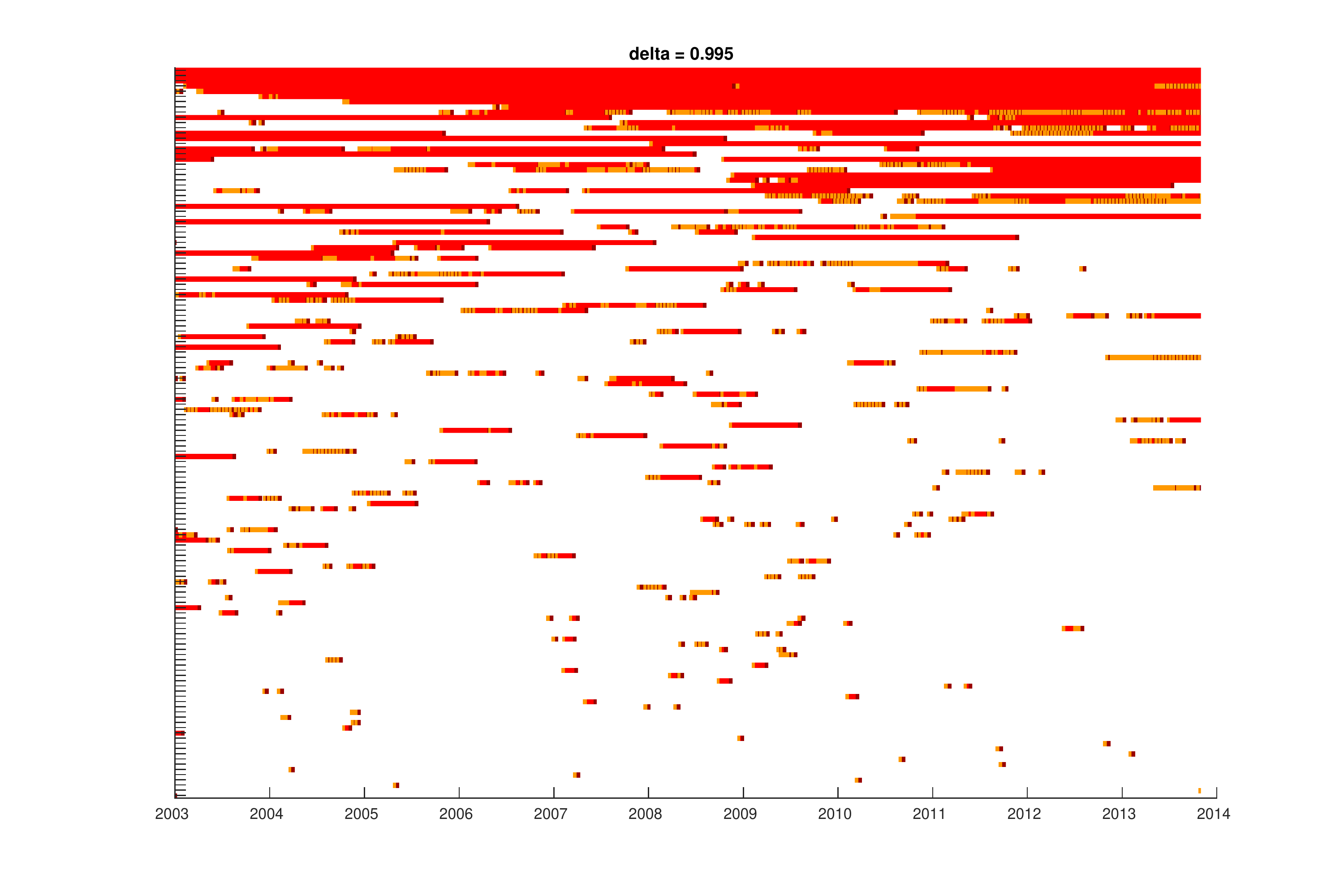} \\
\caption{Evolution of simultaneous parental set of series $j=245$, 3M, in SGDLM  M1.  
The heat-map represents inclusion indicators for the subset of 140 series, of the potential 400, 
that appeared at least once in the warm-up set for 3M. Of these 140 series, 
117 became series became simultaneous parent of 3M at one or more time points. 
Color coding is as follows:   yellow indicates the times of inclusion in the warm-up set,  
red indicates inclusion in the core parental set,  dark red shows inclusion 
in the cool-down set prior to exclusion, and white indicates times when series are not 
included in any of the three sets. 
Series are ordered vertically in decending order of total time
spent in one or more of the three parental sets.} 
\label{sgdlm2:fig:sp-j245-delta995-3MCo-all-sorted}
\end{figure}

\begin{figure}[htbp!]
\centering
\includegraphics[width=7in]{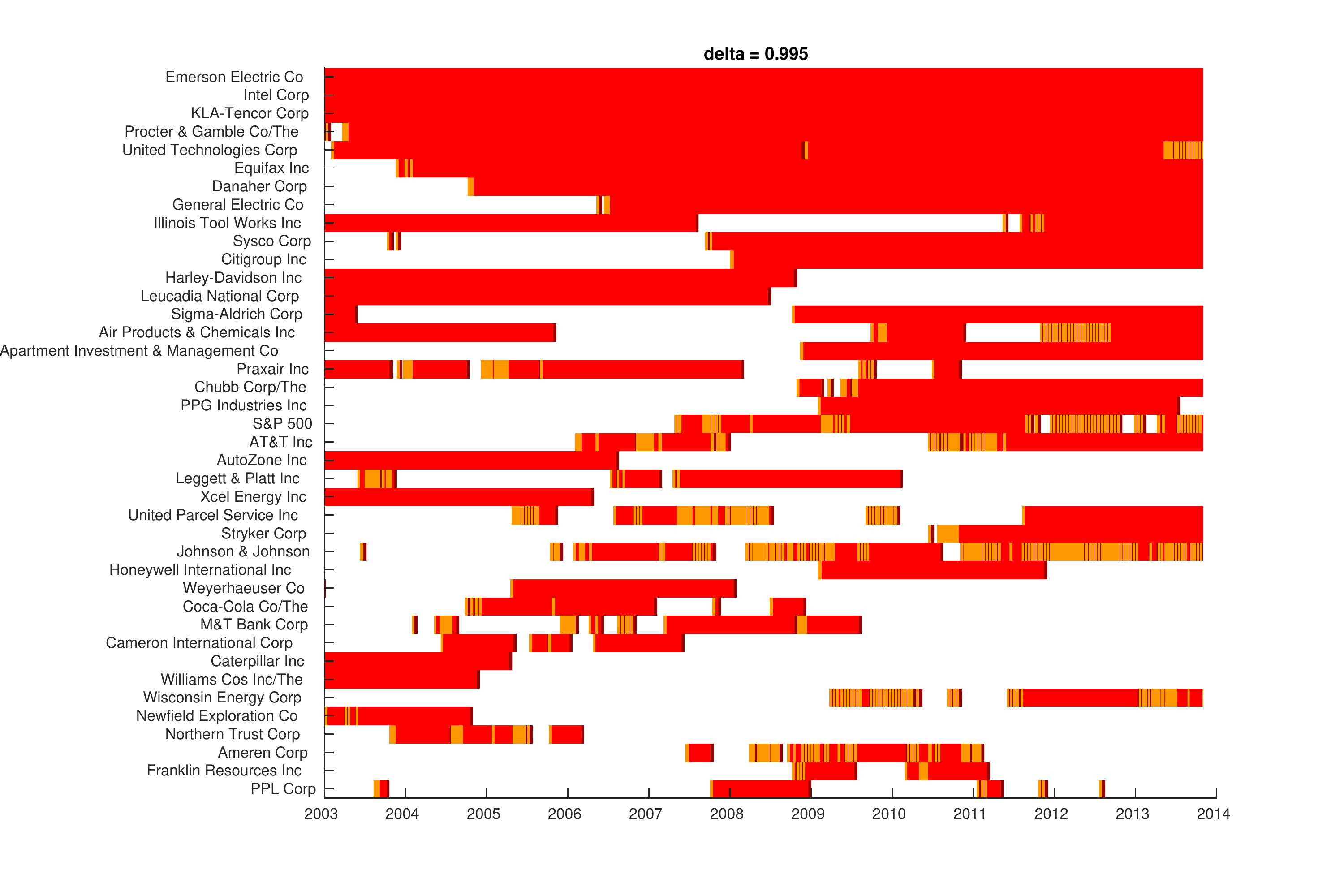} \\
\caption{The first 40 parental series from Figures~\ref{sgdlm2:fig:sp-j245-delta995-3MCo-all-sorted},
i.e., those most active in involvement as candidate or core parental predictors of 3M. 
Note a handful of series are involved as core parents for most of the time period, 
while there is otherwise substantial warm-up/core/exclusion activity over time within this group:
some parents appear only early on, others later, while others
come and go throughout the observation period. } 
\label{sgdlm2:fig:sp-j245-delta995-3MCo-core-top-sorted}
\end{figure}

 Figures~\ref{sgdlm2:fig:sp-j245-delta995-3MCo-all-sorted} and
\ref{sgdlm2:fig:sp-j245-delta995-3MCo-core-top-sorted} display  trajectories indicating 
which series are included as candidate or core parental predictors of returns on 3M stock over time, 
highlighting the adaptive model selection strategy. Tables~\ref{sgdlm2:tb:data-example:investment-returns-all}
and~\ref{sgdlm2:tb:data-example:investment-volas-all} provide additional numerical summaries of 
portfolio performance for all model and portfolio rules evaluated.

\FloatBarrier

\FloatBarrier

\FloatBarrier
\begin{table}[htbp!]
\centering
\small
\begin{tabular}{c c c c c c c c c}
Model & \blue SPX & \red P0 & P1* & P2* & P3* & P4* & P5* & P6* \\  \hline
     & \blue 0.062 & \red 0.091  \\
M1 & & & 0.106 & 0.116 & 0.120 & 0.108 & 0.114 & 0.120 \\
M2 & & & 0.117 & 0.121 & 0.113 & 0.119 & 0.124 & 0.118 \\
M3 & & & 0.092 & 0.086 & 0.089 & 0.089 & 0.087 & 0.089 \\
M4 & & & 0.101 & 0.103 & 0.098 & 0.107 & 0.106 & 0.102 \\
M5 & & & 0.110 & 0.109 & 0.105 & 0.101 & 0.098 & 0.094 \\
MA1& & & 0.093 & 0.086 & 0.082 & 0.094 & 0.091 & 0.088 \\
MA2& & & 0.094 & 0.085 & 0.078 & 0.095 & 0.085 & 0.078 \\
MA3& & & 0.096 & 0.082 & 0.071 & 0.096 & 0.080 & 0.072 \\
MA4& & & 0.102 & 0.092 & 0.088 & 0.104 & 0.093 & 0.089 \\
MA5& & & 0.110 & 0.101 & 0.074 & 0.116 & 0.103 & 0.082 \\
W1 & & & 0.006 &-0.001 &-0.012 & 0.010 & 0.010 &-0.010 \\
W2 & & & 0.006 &-0.000 &-0.011 & 0.010 & 0.002 &-0.009 \\
W3 & & & 0.006 & 0.001 &-0.008 & 0.009 & 0.003 &-0.006 \\
W4 & & & 0.005 & 0.004 &-0.004 & 0.009 & 0.006 &-0.001 \\
W5 & & & 0.005 & 0.006 & 0.003 & 0.008 & 0.009 & 0.006 \\
\end{tabular}
\normalsize
\caption{Annualized log-returns of portfolios SPX, P0, P1*-P6* based on forecasts from models M1-M5, MA1-MA5 and W1-W5 over 2003--2013 after accounting for trading costs.} \label{sgdlm2:tb:data-example:investment-returns-all}
%
\bigskip\bigskip
\centering
\small
\begin{tabular}{c c c c c c c c c}
Model & \blue SPX & \red P0 & P1* & P2* & P3* & P4* & P5* & P6* \\  \hline
        & \blue 0.199 & \red 0.219 \\
M1 & & & 0.150 & 0.147 & 0.140 & 0.153 & 0.150 & 0.143 \\
M2 & & & 0.142 & 0.139 & 0.131 & 0.147 & 0.142 & 0.133 \\
M3 & & & 0.146 & 0.139 & 0.129 & 0.149 & 0.140 & 0.130 \\
M4 & & & 0.149 & 0.151 & 0.146 & 0.152 & 0.154 & 0.148 \\
M5 & & & 0.152 & 0.146 & 0.145 & 0.148 & 0.142 & 0.141 \\
MA1& & & 0.129 & 0.121 & 0.118 & 0.133 & 0.123 & 0.120 \\
MA2& & & 0.121 & 0.115 & 0.110 & 0.124 & 0.118 & 0.114 \\
MA3& & & 0.120 & 0.109 & 0.104 & 0.123 & 0.111 & 0.105 \\
MA4& & & 0.123 & 0.117 & 0.117 & 0.125 & 0.119 & 0.119 \\
MA5& & & 0.125 & 0.117 & 0.107 & 0.128 & 0.117 & 0.107 \\
W1 & & & 0.103 & 0.103 & 0.105 & 0.104 & 0.105 & 0.106 \\
W2 & & & 0.103 & 0.103 & 0.105 & 0.104 & 0.105 & 0.106 \\
W3 & & & 0.103 & 0.103 & 0.105 & 0.104 & 0.105 & 0.106 \\
W4 & & & 0.103 & 0.103 & 0.105 & 0.104 & 0.104 & 0.106 \\
W5 & & & 0.103 & 0.103 & 0.105 & 0.104 & 0.104 & 0.106 \\
\end{tabular}
\normalsize
\caption{Annualized volatilities of portfolios SPX, P0, P1*-P6* based on forecasts from models M1-M5, MA1-MA5 and W1-W5 over 2003--2013 after accounting for trading costs.} \label{sgdlm2:tb:data-example:investment-volas-all}
\end{table}

\FloatBarrier

\newpage
%
 

\begin{thebibliography}{31}
\expandafter\ifx\csname natexlab\endcsname\relax\def\natexlab#1{#1}\fi
\expandafter\ifx\csname url\endcsname\relax
  \def\url#1{\texttt{#1}}\fi
\expandafter\ifx\csname urlprefix\endcsname\relax\def\urlprefix{URL }\fi

\bibitem[{Aguilar and West(2000)}]{AguilarWest2000}
Aguilar, O., West, M., 2000. Bayesian dynamic factor models and portfolio
  allocation. Journal of Business \& Economic Statistics 18~(3), 338--357.

\bibitem[{{Basel Committee on Banking Supervision}(2004)}]{BASEL2}
{Basel Committee on Banking Supervision}, 2004. {International Convergence of
  Capital Measurement and Capital Standards, A Revised Framework}. Bank for
  International Settlements.
\newline\urlprefix\url{http://www.bis.org/publ/bcbs107.pdf}

\bibitem[{Carvalho et~al.(2011)Carvalho, Lopes, and
  Aguilar}]{CarvalhoLopesAguilar2011}
Carvalho, C.~M., Lopes, H.~F., Aguilar, O., 2011. Dynamic stock selection
  strategies: {A} structured factor model framework (with discussion). In:
  Bernardo, J.~M., Bayarri, M.~J., Berger, J.~O., Dawid, A.~P., Heckerman, D.,
  Smith, A. F.~M., West, M. (Eds.), Bayesian Statistics 9. Oxford University
  Press, pp. 69--90.

\bibitem[{Carvalho and West(2007)}]{CarvalhoWest2007}
Carvalho, C.~M., West, M., 2007. Dynamic matrix-variate graphical models.
  Bayesian Analysis 2~(1), 69--98.

\bibitem[{Chan et~al.(2005)Chan, Kohn, and Kirby}]{chankohnkirby2005}
Chan, D., Kohn, R., Kirby, C., 2005. {Multivariate stochastic volatility models
  with correlated errors}. Econometric Reviews 25~(2-3), 245--274.

\bibitem[{Chib et~al.(2006)Chib, Nardari, and
  Shephard}]{chibnardarishephard2006}
Chib, S., Nardari, F., Shephard, N., 2006. {Analysis of high dimensional
  multivariate stochastic volatility models}. Journal of Econometrics 134~(2),
  341--371.

\bibitem[{{Federal Reserve Bank of St. Louis}(2014)}]{stlouisfed-stlfsi}
{Federal Reserve Bank of St. Louis}, 2014. {What is the {S}t. {L}ouis {F}ed
  financial stress index?}
\newline\urlprefix\url{https://www.stlouisfed.org}

\bibitem[{Gruber and Czado(2015)}]{gruberczado2015a}
Gruber, L.~F., Czado, C., 2015. Sequential {B}ayesian model selection of
  regular vine copulas. Bayesian Analysis 10~(4), 937--963.

\bibitem[{Gruber and West(2016)}]{gruberwest2015a}
Gruber, L.~F., West, M., 2016. {GPU}-accelerated {B}ayesian learning in
  simultaneous graphical dynamic linear models. Bayesian Analysis 11, 125--149,
  (advance publication: 2 March 2015).

\bibitem[{Harvey et~al.(1994)Harvey, Ruiz, and
  Shephard}]{harveyruizshephard1994}
Harvey, A.~C., Ruiz, E., Shephard, N., 1994. {Multivariate stochastic variance
  models}. Review of Economic Studies 61, 247--264.

\bibitem[{Jaakkola and Jordan(2000)}]{JaakolaJordan2000}
Jaakkola, T.~S., Jordan, M.~I., 2000. Bayesian parameter estimation via
  variational methods. Statistics and Computing 10, 25--27.

\bibitem[{Jacquier et~al.(2004)Jacquier, Polson, and
  Rossi}]{jacquierpolsonrossi2004}
Jacquier, E., Polson, N.~G., Rossi, P.~E., 2004. {Bayesian analysis of
  stochastic volatility models with fat-tails and correlated errors}. Journal
  of Econometrics 122, 185--212.

\bibitem[{Jones et~al.(2005)Jones, Dobra, Carvalho, Hans, Carter, and
  West}]{JonesDobraCarvalhoHansCarterWest2005}
Jones, B., Dobra, A., Carvalho, C.~M., Hans, C., Carter, C., West, M., 2005.
  Experiments in stochastic computation for high-dimensional graphical models.
  Statistical Science 20, 388--400.

\bibitem[{Kliesen and Smith(2010)}]{kliesensmith2010}
Kliesen, K.~L., Smith, D.~C., January 2010. Measuring financial market stress:
  {T}he {S}t. {L}ouis {F}ed's financial stress index {(STLFSI)}. Federal
  Reserve Bank of St. Louis National Economic Trends.

\bibitem[{Lopes et~al.(2012)Lopes, McCulloch, and Tsay}]{LopesMcCullochTsay12}
Lopes, H.~F., McCulloch, R.~E., Tsay, R., 2012. Cholesky stochastic volatility
  models for high-dimensional time series. Tech. rep., University of Chicago,
  Booth School of Business.

\bibitem[{Markowitz(1952)}]{markowitz1952}
Markowitz, H., 1952. Portfolio selection. The Journal of Finance 7~(1), 77--91.

\bibitem[{Markowitz(1959)}]{markowitz1959}
Markowitz, H., 1959. Portfolio Selection: Efficient Diversification of
  Investments. John Wiley \& Sons, and Chapman \& Hall.

\bibitem[{Nakajima and West(2013)}]{NakajimaWest2010}
Nakajima, J., West, M., 2013. Bayesian analysis of latent threshold dynamic
  models. Journal of Business \& Economic Statistics 31, 151--164.

\bibitem[{Pitt and Shephard(1999)}]{PittShephard1999}
Pitt, M., Shephard, N., 1999. Time varying covariances: {A} factor stochastic
  volatility approach (with discussion). In: Bernardo, J.~M., Berger, J.~O.,
  Dawid, A.~P., Smith, A. F.~M. (Eds.), Bayesian Statistics VI. Oxford
  University Press, pp. 547--570.

\bibitem[{Prado and West(2010)}]{PradoWest2010}
Prado, R., West, M., 2010. {Time Series: Modeling, Computation \& Inference}.
  Chapman \& Hall/CRC Press.

\bibitem[{Quintana et~al.(2010)Quintana, Carvalho, Scott, and
  Costigliola}]{QuintanaCarvalhoScottCostigliola2010}
Quintana, J.~M., Carvalho, C.~M., Scott, J., Costigliola, T., 2010. {Futures
  markets, Bayesian forecasting and risk modeling}. In: O'Hagan, A., West, M.
  (Eds.), {The Handbook of Applied Bayesian Analysis}. Oxford University Press,
  pp. 343--365.

\bibitem[{Quintana et~al.(2003)Quintana, Lourdes, Aguilar, and
  Liu}]{QuintanaLourdesAguilarLiu2003}
Quintana, J.~M., Lourdes, V., Aguilar, O., Liu, J., 2003. Global gambling. In:
  Bernardo, J.~M., Bayarri, M.~J., Berger, J.~O., Dawid, A.~P., Heckerman, D.,
  Smith, A. F.~M., West, M. (Eds.), Bayesian Statistics 7. Oxford University
  Press, pp. 349--368.

\bibitem[{Quintana and West(1987)}]{quintanawest1987}
Quintana, J.~M., West, M., 1987. {An analysis of international exchange rates
  using multivariate DLMs}. The Statistician 36, 275--281.

\bibitem[{Wand et~al.(2011)Wand, Ormerod, Padoan, and
  Fuhrwirth}]{WandOrmerodPadoanFurhwirth2011}
Wand, M.~P., Ormerod, J.~T., Padoan, S.~A., Fuhrwirth, R., 2011. Mean field
  variational {B}ayes for elaborate distributions. Bayesian Analysis 6,
  847--900.

\bibitem[{Wang(2010)}]{Wang2010}
Wang, H., 2010. Sparse seemingly unrelated regression modelling: {A}pplications
  in finance and econometrics. Computational Statistics \& Data Analysis 54,
  2866--2877.

\bibitem[{Wang and West(2009)}]{WangWest2009}
Wang, H., West, M., 2009. Bayesian analysis of matrix normal graphical models.
  Biometrika 96, 821--834.

\bibitem[{West(2003)}]{West2003}
West, M., 2003. Bayesian factor regression models in the ``large p, small n''
  paradigm. In: Bernardo, J.~M., Bayarri, M.~J., Berger, J.~O., Dawid, A.~P.,
  Heckerman, D., Smith, A. F.~M., West, M. (Eds.), Bayesian Statistics 7.
  Oxford University Press, pp. 723--732.

\bibitem[{West and Harrison(1997)}]{WestHarrison1997}
West, M., Harrison, J., 1997. {Bayesian Forecasting \& Dynamic Models}, 2nd
  Edition. Springer Verlag.

\bibitem[{Yoshida and West(2010)}]{YoshidaWest2010}
Yoshida, R., West, M., 2010. Bayesian learning in sparse graphical factor
  models via annealed entropy. Journal of Machine Learning Research 11,
  1771--1798.

\bibitem[{Zhao et~al.(2016)Zhao, Xie, and West}]{ZhaoXieWest2015}
Zhao, Z.~Y., Xie, M., West, M., 2016. Dynamic dependence networks: {F}inancial
  time series forecasting \& portfolio decisions (with discussion). Applied
  Stochastic Models in Business and Industry First published online: March 25,
  2016, --.

\bibitem[{Zhou et~al.(2014)Zhou, Nakajima, and West}]{ZhouNakajimaWest2012}
Zhou, X., Nakajima, J., West, M., 2014. Bayesian forecasting and portfolio
  decisions using dynamic dependent sparse factor models. International Journal
  of Forecasting 30~(4), 963--980.

\end{thebibliography}

\section*{Acknowledgments}
Research presented here was partially developed while the first author was a Visiting Scholar in the Department of Statistical Science at Duke University.  Partial financial support was provided 
by the Fulbright Foundation through the Fulbright Program for Foreign Students (L.F.G.). All opinions, findings and conclusions or recommendations expressed in this work are those of the authors and do not necessarily reflect the views of the Fulbright Foundation.

\subsection*{Author information} 

{\em Lutz Gruber}
is Senior Analyst at e-commerce analytics firm QuantCo, Germany.  His main research foci are in on-line learning of financial time series, dependence analysis with copulas, and statistical econometric modeling.
Lutz received his MS in Mathematical Finance and Actuarial Science at the Technical University of Munich (TUM) in 2011, followed by his PhD in Statistics at TUM in 2015.
 
{\em Mike West} (\href{http://www.stat.duke.edu/~mw}{www.stat.duke.edu/$\tilde{\phantom{.}}$mw}) is The Arts \& Sciences Professor of Statistics \& Decision Sciences in the 
Department of Statistical Science at Duke University. Mike led  development of the department 
from 1990 to 2001,  has served in the establishment-- and as board member-- of several national research institutes and companies, and is past President of the International Society for Bayesian 
Analysis. Mike works in theory and applications of Bayesian statistics, with
highlights in dynamic modeling, time series analysis and forecasting. 
Mike has advised nearly 60 PhD students and postdoctoral associates, and numerous undergraduate and MS students.

\end{document}